\documentclass[conference]{IEEEtran}

\ifCLASSOPTIONcompsoc
  \usepackage[nocompress]{cite}
\else
  \usepackage{cite}
\fi

\usepackage{cite}
\usepackage{amssymb,amsfonts}
\usepackage{graphicx}
\usepackage{textcomp}
\usepackage{color}
\usepackage{algorithm}
\usepackage{algpseudocode}
\floatname{algorithm}{Protocol}

\usepackage{xspace}
\usepackage{amsmath}

\newcommand*{\QEDB}{\hfill\ensuremath{\square}}

\newcommand{\sysname}{\textsc{FedXGB}\xspace}
\newcommand{\othersys}[1]{\textsc{#1}\xspace}
\newcommand{\protocol}[1]{{\textsf{\small{#1}}}}
\newcommand{\func}[1]{{\texttt{{#1}}}}
\newcommand{\makeAngle}[1]{\langle #1 \rangle}

\usepackage{multirow}
\usepackage{array}
\usepackage{diagbox}

\usepackage{subfigure}
\usepackage{epstopdf}

\usepackage{fancyhdr}

\begin{document}
\title{Boosting Privately: Federated Extreme Gradient Boosting for Mobile Crowdsensing}
\author{\IEEEauthorblockN{Yang Liu$^1$, Zhuo Ma*$^1$, Ximeng Liu*$^2$, Siqi Ma$^3$, Surya Nepal$^3$, Robert.H Deng$^4$, Kui Ren$^5$}
\IEEEauthorblockA{$^1$ Xidian University,
  $^2$ Fuzhou University,
  $^3$ Data 61, CSIRO,
  $^4$ Singapore Management University,
  $^5$ Zhejiang University \\
bcds2018@foxmail.com, mazhuo@mail.xidian.edu.cn, snbnix@gmail.com,\\  siqi.ma@csiro.au, surya.nepal@csiro.au, robertdeng@smu.edu.sg, kuiren@zju.edu.cn\\
* Corresponding Author}}

\maketitle
\thispagestyle{fancy} 
\lhead{} 
\chead{} 
\rhead{} 
\lfoot{} 
\cfoot{} 
\rhead{\thepage} 
\renewcommand{\headrulewidth}{0pt} 
\renewcommand{\footrulewidth}{0pt} 
\pagestyle{fancy}
\rfoot{\thepage}

\begin{abstract}
Recently, Google and other 24 institutions proposed a series of open challenges towards federated learning (FL), which include application expansion and homomorphic encryption (HE).
The former aims to expand the applicable machine learning models of FL. 
The latter focuses on who holds the secret key when applying HE to FL.
For the naive HE scheme, the server is set to master the secret key.
Such a setting causes a serious problem that if the server does not conduct aggregation before decryption, a chance is left for the server to access the user's update.
Inspired by the two challenges, we propose \sysname, a federated extreme gradient boosting (XGBoost) scheme supporting forced aggregation.
\sysname mainly achieves the following two breakthroughs.
First, \sysname involves a new HE based secure aggregation scheme for FL.
By combining the advantages of secret sharing and homomorphic encryption, the algorithm can solve the second challenge mentioned above, and is robust to the user dropout. 
Then, \sysname extends FL to a new machine learning model by applying the secure aggregation scheme to the classification and regression tree building of XGBoost.
Moreover, we conduct a comprehensive theoretical analysis and extensive experiments to evaluate the security, effectiveness, and efficiency of \sysname.
The results indicate that \sysname achieves less than 1\% accuracy loss compared with the original XGBoost, and can provide about 23.9\% runtime and 33.3\% communication reduction for HE based model update aggregation of FL.

\end{abstract}


\section{Introduction}
\label{sec_introduction}
Extreme gradient boosting (XGBoost) is a state-of-the-art machine learning model that performs well in processing both classification and regression tasks.
Winning 17 out of 29 challenges published by the world-famous Kaggle competition validates that XGBoost is sure to have an impressive prospect for the further development of artificial intelligence~\cite{chen2016xgboost}.
Similar to other machine learning models, the performance of XGBoost depends on how well the training dataset performs.
However, creating a large dataset requires lots of human efforts, which is an unaffordable cost for most companies.
Hence, mobile crowdsensing is designed to collect data from mobile users who are willing to share data.
DroidNet \cite{rustgi2019droidnet} is a sample system to demonstrate how mobile crowdsensing is used in machine learning model training. 
However, the past crowdsensing architecture usually allows the central server to access to the plaintext user's data, which leaves a chance for user privacy leakage.
The incident of Facebook-Cambridge Analytica happened in 2018, is a significant example to demonstrate the consequences of such privacy leakage.
The large IT company secretly harvested millions of private user data and use them to control a country's leadership election~\cite{garcia2018analyzing}.

To address the privacy leakage problem for mobile crowdsensing, federated learning is proposed by Google and rapidly attract exploded interests of researchers~\cite{mcmahan2016communication}.
Federated learning groups mobile users and the central server into a loose federation, and then, proceeds model training without uploading private user data to the central server.
Despite its excellent features on security and performance, federated learning is still a developing technique.
In 2019, Google, in conjunction with 24 other agencies, proposes a series of open challenges for the future development of federated learning~\cite{kairouz2019advances}.
Besides expanding the applicable machine learning model for federated learning, Google points out that homomorphic encryption (HE) can be a powerful tool in federated learning.
However, existing HE based FL schemes~\cite{yang2019quas, cheng2019secureboost, truex2018hybrid} still have the following two unresolved challenges.
\begin{itemize}
    \item \textbf{Forced Aggregation.}
    The native applications of HE to federated learning is that the server encrypts the parameters with its own public key and sends them to the user~\cite{yang2019quas, cheng2019secureboost}.
    Utilizing the homomorphism of HE, the user can compute the model update without decryption and return the encrypted model update to the server for aggregation.
    A key challenge here is to force aggregation on the server before decryption, as otherwise, the server may be able to learn a user’s model update.

    \item \textbf{User Dropout.}
    Federated learning is originally designed to run in the open network, in which the user's connectivity is always unstable.
    To date, most of the existing HE based federated learning schemes cannot resolve the accident user dropout problem, other than abandoning the current round of training~\cite{yang2019quas, cheng2019secureboost, truex2018hybrid}.
    Such a drawback dramatically reduces the practicality of the schemes in applications.
\end{itemize}

To resolve the above challenges, we propose \sysname, a federated extreme gradient boosting framework for mobile crowdsensing that supports forced aggregation, and is robust against user dropout. 
\sysname is composed of two kinds of entities, a central cloud server and a set of users.
\sysname proceeds as follows.
The central server iteratively invokes a suite of secure protocols to build the classification and regression tree (CART) of XGBoost.
In the protocols, our newly designed secure aggregation protocol is invoked to aggregate the users' gradients.
By combining Bresson's cryptosystem~\cite{bresson2003simple} and Shamir's secret sharing~\cite{shamir1979share}, \sysname makes the central server to operate a forced aggregation on the gradients and can recover the data of the dropout users.

Our contributions can be summarised as follows:
\begin{itemize}
    \item \textbf{Federated Extreme Gradient Boosting.}
    We propose a federated learning framework to implement privacy-preserving XGBoost training for mobile crowdsensing, called \sysname.
    Using a suite of secure protocols, \sysname allows multiple users to cooperatively train an XGBoost model without direct revealing of their private data to the central server.
    
    \item \textbf{Forced Aggregation.}
    We design a new secure gradient aggregation algorithm for federated learning, which combines the advantages of both homomorphic encryption and secret sharing.
    Specifically, through the combination of homomorphic encryption and secret sharing, \sysname ensures that the central server cannot get a correct decryption result before operating aggregation, and meanwhile, is robust against user dropout.
    
    \item \textbf{Practicality for Applications.}
    We evaluate the effectiveness and efficiency of \sysname using two standard datasets. 
    The results indicate that \sysname maintains the high performance of XGBoost with less than 1\% of accuracy loss and attains about 23.9\% runtime and 33.3\% communication reduction for gradient aggregation.
\end{itemize}

\noindent
\textbf{Outline.}
The rest of this paper is organized as follows. 
In Section \ref{sec_preminary}, some background knowledge are briefly introduced. 
Section \ref{sec_overview} gives an overview of \sysname. 
Section~\ref{sec_technical} presents the technical intuition of our secure aggregation scheme.
Section \ref{sec_approach} lists the implementation details of \sysname. 
The security and performance of \sysname are discussed and evaluated in Section \ref{sec_analysis} and Section \ref{sec_experiments}. 
Section \ref{sec_relatedwork} discusses the related work. 
The last section concludes the paper.

\section{PRELIMINARY}
\label{sec_preminary}
In this section, we briefly introduce the background knowledge about XGBoost and the cryptographic functions used in \sysname.
Table~\ref{table_notation} summarizes the frequently-used notations. 
\vspace{-0.6cm}
\renewcommand\arraystretch{1.3}
\begin{table}[!htbp]
\centering
\caption{Notation Table}
\begin{tabular}{cl}

\hline
Notation & Description       \\

\hline

$w_i$               &  the first-order derivative of $l(\cdot)$ for the $i_{th}$ instance. \\

$h_i$               &  the second-order derivative of $l(\cdot)$ for the $i_{th}$ instance. \\


$\zeta_{u, v}$      & the secret share distributed to the user $u$ by the user $v$. \\

$\mathbb{F}$       & a finite field $\mathbb{F}$, e.g. $\mathbb{F}_p=\mathbb{Z}*_p$ for some large prime $p$. \\

$\mathbb{G}$        & a cyclic group with a generator $g$.\\

$\langle\cdot\rangle_u$        & key used for secure aggregation.\\

$[\![x]\!]$        & an encrypted secret $x$.\\

\hline
\end{tabular}
\label{table_notation}
\end{table}
\vspace{-0.2cm}

\subsection{Extreme Gradient Boosting}\label{sub_prexgboost}
One of the goals of \sysname is to extend federated learning to the ensemble learning model XGBoost.
An XGBoost model is composed of multiple classification and regression trees (CARTs) that are built based on the boosting method~\cite{chen2016xgboost}.
For the $k$-th iteration,
    the objective of XGBoost is to generate a CART to minimize the following objective function $\mathcal{L}_{k}$.
\begin{equation}\label{eq_objective}
\begin{aligned}
    \mathcal{L}_{k} = \sum^n_{i=1} l(y_i, \hat{y}_{k-1, i} + f_k(x_i)) + \Omega (f_k),\\
\end{aligned}
\end{equation}
where $n$ is the total number of training samples, 
    $i$ is the index of each sample,
    and $y_i$ is the label of the $i$-th sample,
    $\hat{y}_{k-1, i}$ represents the predicted label of the $i$-th sample at the ($k - 1$)-th iteration,
    $\Omega$ is a regularization item.
To grow a CART, XGBoost iteratively adds branches (i.e., splitting the leaf node) to the current tree.
Assume $I_L$ and $I_R$ are the instance sets of left and right nodes after a split, and $I = I_L\cup I_R$.
The score to evaluate a split is as follows.
\begin{equation}\label{eq_score}
    score = \frac{1}{2}\cdot (\frac{(\sum_{i\in I_L} w_i)^2}{\sum_{i\in I_L} h_i + \lambda} + \frac{\sum_{i\in I_R} w_i^2}{\sum_{i\in I_R} h_i + \lambda} - \frac{(\sum_{i\in I} w_i)^2}{\sum_{i\in I} h_i + \lambda}),
\end{equation}
where $\lambda$ is a constant value, $w_i$ and $h_i$ are the first-order and second-order derivatives of $l(\cdot)$.
Each time a branch is added, XGBoost chooses the split with the maximum score from all candidate splits.
When a CART structure is fixed,
    the weight $\omega_j$ of a leaf node $j$ is calculated by Eq.~\ref{eq_weight}.
\begin{equation}\label{eq_weight}
    \omega_j = - \frac{(\sum_{i\in I} w_i)^2}{\sum_{i\in I} h_i + \lambda}.
\end{equation}

\subsection{Homomorphic Encryption}
Bresson's cryptosystem~\cite{bresson2003simple} is a kind of partially homomorphic cryptosystem derived from Paillier's cryptosystem.
\sysname adopts it to protect the gradient of users.
The followings are some definitions of the cryptosystem.

\textbf{Key Generation.}
Three inputs are required for the key generation function \func{Key.Gen}, namely a security parameter $\ell$, a big integer $N = q_1 q_2$ and a generator $g$.
$q_1$ and $q_2$ are two primes that satisfy $q_1 = 2q_1' + 1$ and $q_2 = 2q_2' + 1$, where $q_1'$ and $q_2'$ are primes and different from $q_1$ and $q_2$.
$g\in \mathbb{Z}_{N_2}^*$ is a generator of the group $(\mathbb{G}, q_1, q_2, N = q_1 q_2, g)$ with order $ord(\mathbb{G}) = (p - 1)(q - 1)/2$.
With the inputs, \func{Key.Gen} outputs a private-public key pair $(\makeAngle{k_{pri}}, \makeAngle{k_{pub}})$, where $\makeAngle{k_{pri}}\in [1, ord(\mathbb{G}))$ and $\makeAngle{k_{pub}} = g^{\makeAngle{k_{pri}}}\mod N^2$.

\textbf{Encryption \& Decryption.}
To encrypt a message $m\in \mathbb{Z}_N^*$, a random value $r\in \mathbb{Z}_{N}^*$ is first chosen.
Then, using the public key $\makeAngle{k_{pub}}$, we can compute the ciphertext $(c_1, c_2)$ according to Eq.~\ref{eq_hoenc}
\begin{equation}\label{eq_hoenc}
  c_1 = g^r\mod N^2, c_2 = (1 + mN)\makeAngle{k_{pub}}^r \mod N^2.
\end{equation}
Knowing the private key, we can decrypt the ciphertext as follows.
\begin{equation}
  m =  \frac{1}{N}(c_2 / c_1^{\makeAngle{k_{pri}}} - 1 \mod N^2).  
\end{equation}
The above encryption is semantically secure under the decisional Diffie-Hellman assumption in $\mathbb{Z}^*_{N^2}$~\cite{bresson2003simple}.

\textbf{Key Agreement.}
\sysname invokes the key agreement function \func{Key.Agr} to generate the shared key used for the shared key encryption and secret masking in the secure aggregation process.
Given a user's private key $\makeAngle{k_{pri, u}}$ and a public key of another user $\makeAngle{k_{pub, v}}$, \func{Key.Agr} outputs $\makeAngle{k_{u, v}} = \makeAngle{k_{pub, v}}^{\makeAngle{k_{pri, u}}}\mod N^2$.
It can be proved that \func{Key.Agr} is also secure under the decisional Diffie-Hellman assumption in $\mathbb{Z}^*_{N^2}$~\cite{bresson2003simple}.

\subsection{Secret Sharing}
\sysname utilizes Shamir's secret sharing scheme~\cite{shamir1979share} to deal with the user dropout problem.
Two functions are involved in the Shamir's secret sharing scheme, share generation \func{SS.Share} and secret reconstruction \func{SS.Recon}.
Given the secret $s$, the threshold $t$ and a set of users $\mathcal{U}$, \func{SS.Share}$(s, t, n)$ outputs a set of shares for each user $\{(u, \zeta_u)| u\in \mathcal{U}\}$.
Inversely, given at least $t$ shares, \func{SS.Recon} recovers the secret $s$ by the Lagrange polynomials.
For security, \func{SS.Share} and \func{SS.Recon} work on a finite field $\mathbb{F} = \mathbb{Z}^*_p$, where $p$ is a big prime.
Note that since \sysname uses the above functions to secretly share the private key, $p$ has to be bigger than $ord(\mathbb{G})$, which is the order of the group used in \func{Key.Gen}.




\subsection{Share-Key Encryption}
\sysname uses the shared-key encryption to avoid the adversary's eavesdropping while transmitting the secret shares.
For security, the encryption function \func{Enc} is required to be indistinguishable under a chosen plaintext attack (IND-CPA).
Given a share key $\makeAngle{k_{u, v}}$ and message $m$, \func{Enc} outputs a ciphertext $c_{u, v} = $\func{Enc}$(\makeAngle{k_{u, v}}, m)$, and \func{Dec}$(\makeAngle{k_{u, v}}, c_{u, v})$ outputs the plaintext $m$.

\section{Overview of \sysname}
\label{sec_overview}
\begin{figure}[ht!]
\centering
\includegraphics[scale=0.75]{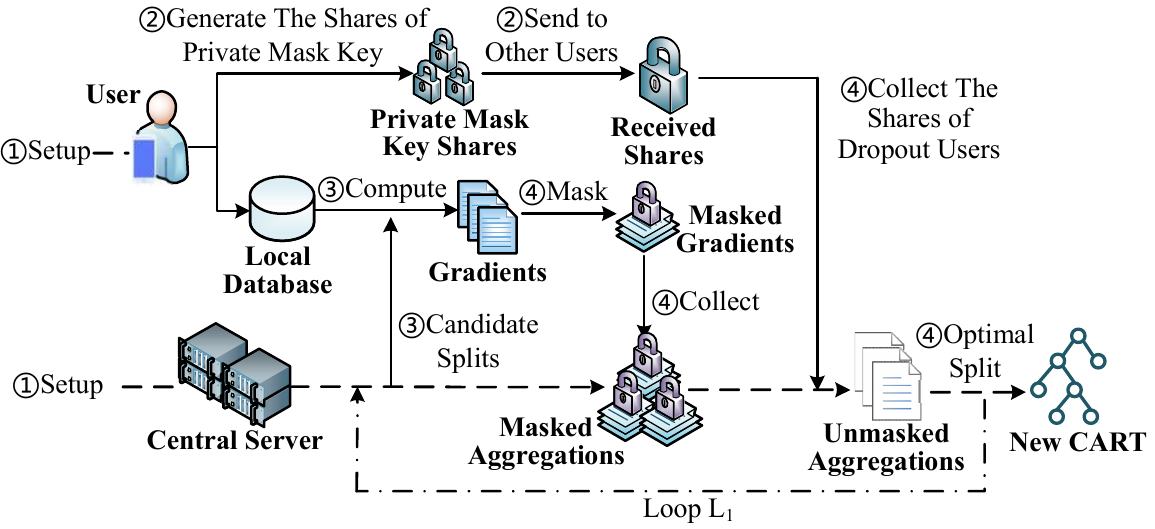}
\caption{Entities of \sysname}
\label{fig_system_model}
\end{figure}

In this section, we briefly overview the system design of \sysname for mobile crowdsensing.
To provide a better understanding of \sysname, we first list the entities of \sysname, illustrated in Fig.~\ref{fig_system_model}.
Then, we define the security model of \sysname.
Finally, the workflow of \sysname is presented, shown in Fig.~\ref{fig_system_workflow}.

\subsection{Entities of \sysname}
\sysname consists of three types of entities: 
    a set of users $\mathcal{U}$
    and a central server $\mathcal{S}$.

\textbf{Users.} 
$\mathcal{U} = \{u_1, u_2, ..., u_{n}\}$.
Each $u\in \mathcal{U}$ is a mobile user that volunteers to participate in federated learning and is connected with other users and the central server.

\textbf{Central Server.} 
$\mathcal{S}$ is owned by a mobile crowdsensing service provider.
The aggregation of the model update for \sysname proceeds in $\mathcal{S}$, but $\mathcal{S}$ is not trusted by the users.

\subsection{Security Model}
In \sysname,
    our security model is based on the \textit{curious-but-honest} model, 
    a standard security model in federated learning~\cite{mcmahan2016communication, bonawitz2017practical, truex2018hybrid}.
In the model, each entity of the protocol is \textit{curious-but-honest}, defined in \textit{Definition 1}.

\noindent 
\textbf{Definition 1~\cite{paverd2014modelling}.}
\textit{In a communication protocol, a curious-but-honest entity
    does not deviate from the defined protocol, 
    but attempts to learn all possible information from the legitimately received messages.}

In addition, we introduce an active adversary into our security model who has the following abilities.
$\mathcal{A}$ has the following abilities:
1) simultaneously corrupt less than $t$ legitimate users and the central server;
2) eavesdrop the communication channels;
3) for the corrupted entities, $\mathcal{A}$ can access to all their data in plaintext, e.g.,
private keys and random seeds;
4) is limited to have polynomial-time computation power.
\sysname needs to achieve the following security goals.
\begin{itemize}
    \item \textbf{Goal 1: Data Privacy.} 
    $\mathcal{S}$ cannot learn the private data of $u\in \mathcal{U}$, no matter $u$ is active or loses connection in the training process.
    
    \item \textbf{Goal 2: Forced Aggregation.} 
    Considering the secure gradient aggregation of federated learning, we limit that $\mathcal{S}$ cannot ignore a specific user's uploaded model update without prior notice.
\end{itemize}

\subsection{Workflow of \sysname}
Three protocols are involved in \sysname, namely secure CART building (\protocol{SecBoost}), secure split finding (\protocol{SecFind}) and secure aggregation (\protocol{SecAgg}).
\protocol{SecFind} and \protocol{SecAgg} are two sub-protocols of \protocol{SecBoost}.
Here, we briefly overview how the three protocols work in \sysname.
\begin{figure}[ht!]
\centering
\includegraphics[scale=0.65]{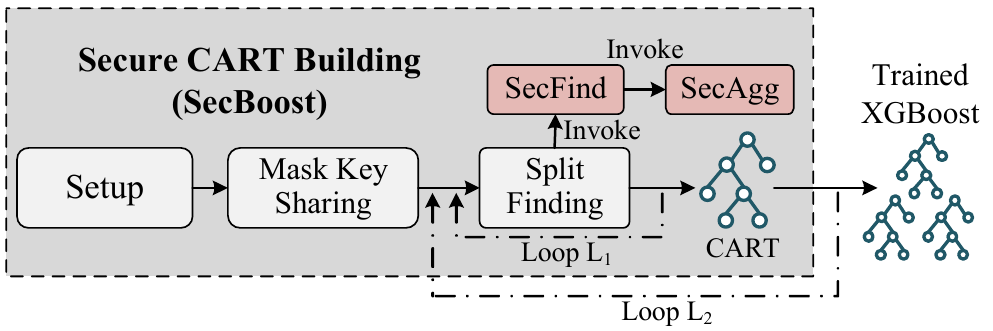}
\caption{Workflow of \sysname}
\label{fig_system_workflow}
\end{figure}

As shown in Fig~\ref{fig_system_workflow}, \sysname trains an XGBoost model by iteratively invokes \protocol{SecBoost}.
\protocol{SecBoost} takes three steps, which are setup, mask key sharing and split finding.
\protocol{SecFind} is used to complete the split finding step of \protocol{SecBoost}, and \protocol{SecAgg} is invoked by \protocol{SecFind} to securely aggregate gradients.
\begin{enumerate}
    \item \textbf{Setup.} 
    A trusted key generate center setups the cryptographic parameters. 
    $\mathcal{U}$ and $\mathcal{S}$ utilize the parameters to generate their cryptographic keys used in the following steps.
    We say that such a key generation center is a typical role in modern networks, e.g., the digital certificate management center.

    \item \textbf{Mask Key Sharing.}
    Before uploading the gradients for CART building, each user secretly shares its private mask key.
    Thus, even the user accidentally drops out, $\mathcal{S}$ can still recover its mask key and continue to get a correct gradient aggregation result.

    \item \textbf{Split Finding.}
    As stated in Section~\ref{sub_prexgboost}, an XGBoost model is composed of multiple CARTs.
    To build a CART, the key operation is to find the optimal split for the leaf node.
    \protocol{SecBoost} achieves split finding by invoking \protocol{SecFind}.
    In \protocol{SecFind}, the user first locally calculates the gradients based on the candidate splits published by $\mathcal{S}$.
    The gradients include both $w_i$ and $h_i$, defined in Eq.~\ref{eq_score}.
    Then, $\mathcal{S}$ aggregates each user's gradients by \protocol{SecAgg}. 
    Having the aggregation result, $\mathcal{S}$ can derive the scores of all candidate splits according to Eq.~\ref{eq_score}.
    Finally, $\mathcal{S}$ chooses the one with the maximum score and use it to add a new branch in the current CART.
\end{enumerate}

By repeating the above steps (Loop L$_1$ in Fig.~\ref{fig_system_workflow}), $\mathcal{S}$ can obtain a well trained CART $f_k$.
Further, $\mathcal{S}$ publishes the newly trained CART and continues to build the next CART (Loop L$_2$ in Fig.~\ref{fig_system_workflow}). 
In the end, $\mathcal{S}$ gets a trained XGBoost model $ f(x) = \sum_{\kappa  = 1}^K f_{\kappa}(x)$, where $K$ is the maximum training round.

\section{Technical Intuition}
\label{sec_technical}
For federated learning, the most critical operation is the secure gradient aggregation.
In this section, we introduce the intuition to design our secure aggregation protocol and present its implementation details.

\begin{algorithm*}[ht]
  \caption{Secure Aggregation (\protocol{SecAgg})}
  \label{pro_secagg}
  \begin{algorithmic}[1]
    \Require
      A server $\mathcal{S}$;
      a user set $\mathcal{U}$;
      $u\in \mathcal{U}$ holds a secret $x_u$ and a set of other user's private mask shares $\{\zeta_{v, u}| v\in \mathcal{U}/u\}$.
    \Ensure
      $\mathcal{S}$ obtains the aggregated users' secrets $\Lambda$.
    
    \For{$u\in \mathcal{U}$}
        \State Generate $\makeAngle{sk_{u, v}} \gets $\texttt{KEY.Agr}$(\makeAngle{sk_{pri, u}}, \makeAngle{sk_{pub, v}})$ for $v\in \mathcal{U}/u$ and selects a random value $r_u\in \mathbb{Z}_N$.
        
        \State Compute $[\![x_u]\!]\gets$\texttt{SecMask}$(x_u, r_{u}, \makeAngle{sk_{u, v}}, \makeAngle{sk_{pub, \mathcal{S}}})$ and send $[\![x_u]\!]$ to $\mathcal{S}$.
    \EndFor
    
    \State $\mathcal{S}$ checks the dropout users in the above iteration and publishes the active user list $\mathcal{U}'$.
    
    \State Each user checks whether it is in $\mathcal{U}'$. If yes, send $g^{r_u}$ and $\{\zeta_{v, u}| v\in \mathcal{U}/\mathcal{U}'\}$ to $\mathcal{S}$, otherwise, wait for the next invocation.
    
    \State $\mathcal{S}$ computes $\mathcal{R} \gets \prod_{u\in \mathcal{U}'} g^{r_{u}\varphi(g^{r_u})}$, and for $u_0\in \mathcal{U}/\mathcal{U}'$, recovers $\makeAngle{sk_{u_0}^{pri}} \gets $\texttt{SS.Recon}$(\{\zeta_{u_0, v}| v\in \mathcal{U}'\}, t)$ to compute $\{\makeAngle{sk_{u_0, v}}| v\in \mathcal{U}'\}$ and $\Upsilon \gets \sum_{u_0\in \mathcal{U}/ \mathcal{U}'}(\sum_{u_0 < v, v\in \mathcal{U}} \varphi(\makeAngle{sk_{u_0, v}}) -
    \sum_{u_0 < v, v\in \mathcal{U}} \varphi(\makeAngle{sk_{u_0, v}}))$.
    
    \State $\mathcal{S}$ obtains $\sum_{u\in\mathcal{U}'} x_u \gets \frac{1}{N}[(\prod_{u\in \mathcal{U}'} [\![x_u]\!])\cdot
    \mathcal{R}^{-\makeAngle{sk_{pri, \mathcal{S}}}} - 1 + \Upsilon] \mod N^2$.
  \end{algorithmic}
\end{algorithm*}

\subsection{IBM's Homomorphic Aggregation Scheme}
To ensure an \textit{honest-but-curious} central server to reliably aggregate data, a popular method is using the homomorphic encryption technique.
Recently, IBM proposes such a homomorphic aggregation scheme for federated learning~\cite{truex2018hybrid} (abbreviated as \othersys{IBMHom}) as follows.

\othersys{IBMHom} is based on the $t$-threshold Paillier cryptosystem~\cite{damgard2001generalisation}, that is, the ciphertext must be decrypted with more than $t$ secret keys.
Assume that each user $u$ holds a gradient $x_u$ and one of the secret keys.
Express the $t$-threshold encryption algorithm as \func{ThEnc}.
To aggregate all users' gradients, $u$ samples a random value $r_u\in \mathbb{Z}^*_N$ and computes:
\begin{equation}
    [\![x_u]\!] = \func{ThEnc}(x_u + \epsilon_u, r_u, \makeAngle{tk_{pub, \mathcal{S}}}),
\end{equation}
where $\makeAngle{sk_{pub, \mathcal{S}}}$ is the public key and $\epsilon$ is the Gaussian noise to implement differential privacy against the inference attack~\cite{wang2019beyond}.
Then, $[\![x_u]\!]$ is uploaded.
Using the homomorphism of the Paillier cryptosystem, the central server $\mathcal{S}$ can get $[\![\sum_{u\in \mathcal{U}}x_u]\!] = \prod_{u\in \mathcal{U}} [\![x_u]\!]$.
Next, $\mathcal{S}$ randomly chooses $t$ users and orderly ask them to decrypt $[\![\sum_{u\in \mathcal{U}}x_u]\!]$.
In the process, if any user drops out, $\mathcal{S}$ has to ask another user to decrypt.
Finally, with the partial decryption results of all the $t$ users, $\mathcal{S}$ can recover the plaintext aggregation result.

\subsection{Our Hybrid Masking Scheme \protocol{SecAgg}}
We notice that the secure aggregation scheme above has two disadvantages.
First, the aggregation of $\mathcal{S}$ is unforced.
If a malicious server does not aggregate a specific user's data, it can still get a correct decrypted aggregation result of the remaining data.
In other words, a malicious server can directly ask the users to decrypt a specific user's data, and the users cannot be aware of this.
Also, the vulnerability makes it easier to launch an update-leak attack~\cite{salem2019updates}.
Second, the user is responsible for sending, encrypting and decrypting the data at the same time, which is too costly for a user.
To resolve the problems, we propose a new secure aggregation scheme \protocol{SecAgg}, shown in Protocol~\ref{pro_secagg}.

In \protocol{SecAgg}, $u$ first computes the shared mask keys with other users by \func{Key.Agr} and secretly share its private mask key.
Then, $u$ samples a random value $r_u$ and masks $x_u$ using the masking function \func{SecMask}.
\begin{equation}\label{eq_secmask}
\begin{aligned}
    \text{$[\![x_u]\!]$} = &\func{SecMask}(x_u, r_u, \makeAngle{sk_{u, v}}, \makeAngle{sk_{pub, \mathcal{S}}}) \\
        = &[1 + (x_u +\Upsilon_u)N]\cdot \makeAngle{sk_{pub, \mathcal{S}}}^{r_u \varphi(g^{r_u})}  \mod N^2.
\end{aligned}
\end{equation}
where $\Upsilon_u = \sum_{u < v} \varphi(\makeAngle{sk_{u, v}}) - \sum_{u > v} \varphi(\makeAngle{sk_{u, v}})$, $\makeAngle{sk_{pub, \mathcal{S}}}$ is the public mask key of the central server $\mathcal{S}$, $\varphi(\cdot)$ is a pseudo-random function with a fixed-length output.
$[\![x_u]\!]$ is sent to $\mathcal{S}$.
In the process, $\mathcal{S}$ records the received mask values and the senders and then publishes a list of the senders $\mathcal{U}'$.
The active users in $\mathcal{U}'$ return $g^{r_u}$ and the shares of the dropout users' private mask keys. 
Using the shares, $\mathcal{S}$ recovers the private mask key of the dropout users and computes the shared mask keys between the dropout users and the other users.
Finally, the aggregation result can be obtained according to line 8, Protocol~\ref{pro_secagg}.
Notably, it is observed that \protocol{SecAgg} can be simply extended with differential privacy in a similar way of \othersys{IBMHom}.
However, since the inference attack is not the fundamental problem of this paper, we omit the extended implementation.


\vspace{0.1cm}
\textbf{Correctness.}
Express the dropout users as $u_0\in \mathcal{U}/\mathcal{U}'$.
Based on our cryptographic definitions, the unmasking process (line 8, Protocol~\ref{pro_secagg}) can be expressed as Eq.~\ref{eq_correctness_1} and Eq.~\ref{eq_correctness_2}.
\begin{equation}\label{eq_correctness_1}
\begin{aligned}
    \sum_{u\in \mathcal{U}'}x_u
    = &[1 + N \sum_{u\in \mathcal{U}'} (x_u + \Upsilon_u)]\cdot \makeAngle{sk_{pub, \mathcal{S}}}^{\sum_{u\in \mathcal{U}} r_u \varphi(g^{r_u})} \\
    &\cdot g^{-\makeAngle{sk_{pri, \mathcal{S}}} \sum_{u\in \mathcal{U}} r_u \varphi(g^{r_u})}
    +  \sum_{u\in \mathcal{U}/ \mathcal{U}'} \Upsilon_u \mod N^2,
\end{aligned}
\end{equation}
\begin{equation}\label{eq_correctness_2}
\begin{aligned}
    &\begin{cases}
        \Upsilon_u = \sum_{u < v} \varphi(\makeAngle{sk_{u, v}}) - \sum_{u > v} \varphi(\makeAngle{sk_{u, v}}) \mod p\\
        \sum_{u\in \mathcal{U}} \Upsilon_u = \sum_{u\in \mathcal{U}'} \Upsilon_u + \sum_{u\in \mathcal{U}/ \mathcal{U}'} \Upsilon_u =  0
    \end{cases}.
\end{aligned}    
\end{equation}
Since $\makeAngle{sk_{pub, \mathcal{S}}} = g^{\makeAngle{sk_{pri, \mathcal{S}}}}$, we can prove that the unmasking result is the correct aggregation result by combining the above two equations.

\section{Secure CART building of \sysname}
\label{sec_approach}
In this section, we present the details of \sysname for federated XGBoost  training.
In the protocols, the following notions are specified. 
All users are orderly labeled by a sequence of indexes ($1, 2, ..., n$) to represent their identities. 
Each user is deployed with a small local dataset $D_u$.

\subsection{Secure CART Building \protocol{SecBoost}}\label{sub_secboost}
\begin{figure}[ht!]
\centering
\includegraphics[scale=0.7]{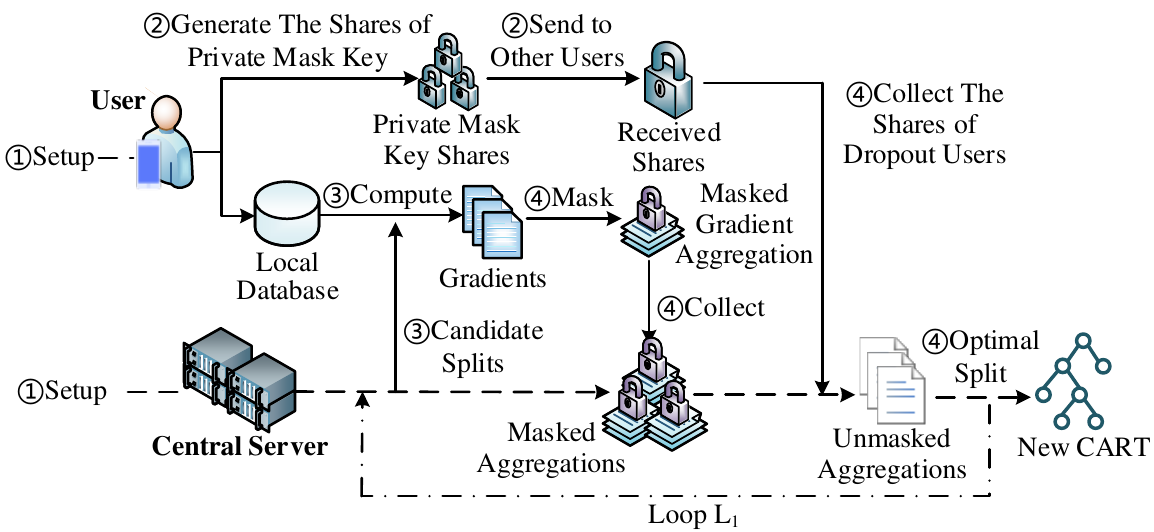}
\caption{Detailed Overview of \protocol{SecBoost}}
\label{overflow}
\end{figure}

As mentioned before, an XGBoost model is composed of multiple CARTs.
\sysname implements secure CART building by invoking \protocol{SecBoost}, shown in Protocol~\ref{FedLearning}.
Referring to the overview illustrated in Fig.~\ref{overflow}, we introduce the detailed steps of \protocol{SecBoost}.

\textbf{Step 1 - Setup:} 
In the step, $\mathcal{U}$ and $\mathcal{S}$ setup the cryptographic keys.
To achieve this, a trusted key generation center samples the parameters for key generation and secret sharing, including a cyclic group $(\mathbb{G}, q_1, q_2, N = q_1 q_2, g)$ and a finite field $\mathbb{Z}^*_p$.
Then, the public cryptographic parameters $(\mathbb{G}, g, N)$ and $\mathbb{Z}^*_p$ are published to both $\mathcal{U}$ and $\mathcal{S}$.
Using the public parameters, each $u\in\mathcal{U}$ invokes \func{Key.Gen} to generate two pairs of keys, $(\makeAngle{ek_{pri,  u}}, \makeAngle{ek_{pub,  u}})$ and $(\makeAngle{sk_{pri,  u}}, \makeAngle{sk_{pub,  u}})$, which are used for shared key encryption and secret masking in secure aggregation, respectively.
Meanwhile, $\mathcal{S}$ generates a pair of masking keys, $(\makeAngle{sk_{pri, \mathcal{S}}}, \makeAngle{sk_{pub, \mathcal{S}}})$ and determines the secret sharing threshold $t$.
Finally, $\mathcal{U}$ and $\mathcal{S}$ exchange their public keys.



\textbf{Step 2 - Mask Key Sharing:} 
To deal with the user's accident dropout, each user previously generates random shares of its private mask keys.
For a specific user $u$, it computes $\{(u, \zeta_{u, v})| v\in$ $ \mathcal{U}\} \gets$ \texttt{SS.Share}$(\makeAngle{sk_{pri,  u}}, t, n)$.
For each selected user $v\in\mathcal{U}$, $u$ encrypts one of the shares $c_{u, v}\gets $\func{Enc}$(\makeAngle{ek_{u, v}},$ $ u||v||\zeta_{u, v})$ and sends the encryption result to it, where $\makeAngle{ek_{u, v}}\gets $\func{Key.Agr}$(\makeAngle{ek_{pri,  u}}, \makeAngle{ek_{pub, v}})$.
The user $v$ decrypts $c_{u, v}$ and extracts $\zeta_{u, v}$.
$\zeta_{u, v}$ is stored and used to recover the private mask key if $u$ drops out.

\textbf{Step 3 - Split Finding:} 
Assume that the feature set of the user data is $\mathcal{Q} = \{\alpha_1, \alpha_2, ..., \alpha_q \}$. 
According to the boosting method defined in XGBoost~\cite{chen2016xgboost}, $\mathcal{S}$ randomly selects a sub-sample of all features $\mathcal{Q}'\subset \mathcal{Q}$ and inputs it to \protocol{SecFind} to find the optimal split.
The detailed split finding method and optimization criteria are stated in the next section.
To build a new CART with an optimal structure, $\mathcal{S}$ successively operates the above steps until the current tree depth reaches the maximum depth or other termination conditions are met~\cite{chen2016xgboost}.
Finally, \protocol{SecBoost} outputs a well trained CART $f_{\kappa}$.

\begin{algorithm*}[ht!]
  \caption{Secure Extreme Gradient Boosting Based Tree Building (\protocol{SecBoost})}
  \label{FedLearning}
  \begin{algorithmic}[1]
    \Require
      A central server $\mathcal{S}$, a user set $\mathcal{U} = \{u_1, ..., u_{n}\}$ and a key generation center $\mathcal{T}$.
    \Ensure
      A well-trained CART.
    
    \State \textbf{Step 1 - Setup:}
    
    \State Given the security parameter $\ell$, $\mathcal{T}$ randomly selects three strong primes $p$, $q_1$ and $q_2$, and samples a cyclic group $(\mathbb{G}, q_1, q_2, N = q_1 q_2, g)$, where $g$ is a generator with order $ord(g) = \frac{(q_1 - 1)(q_2 - 1)}{2}$ and $p > ord(g)$.
    $(\mathbb{G}, g, N)$ and $p$ are published to both $u\in\mathcal{U}$ and $\mathcal{S}$.
    
    \State Each $u\in \mathcal{U}$ compute $(\makeAngle{ek_{pri,  u}}, \makeAngle{ek_{pub,  u}})\gets $\func{Key.Gen}$(g, N, \ell)$ and $(\makeAngle{sk_{pri,  u}}, \makeAngle{sk_{pub,  u}})\gets $\func{Key.Gen}$(g, N, \ell)$.
    
    \State $\mathcal{S}$ generates $(\makeAngle{sk_{pri, \mathcal{S}}}, \makeAngle{sk_{pub, \mathcal{S}}})\gets $\func{Key.Gen}$(g, N, \ell)$ and determines the secret sharing threshold $t$.
    
    \State $\mathcal{S}$ and $\mathcal{U}$ publish their public keys.
    
    
    
    \State \textbf{Step 2 - Mask Key Sharing:}
    
    \State $u\in \mathcal{U}$ computes the shares of its private mask key $\makeAngle{sk_{pri,  u}}$ by $\{(u, \zeta_{u, v})| v\in \mathcal{U}\} \gets$ \texttt{SS.Share}$(\makeAngle{sk_{pri,  u}}, t, n)$.
    
    \State For $v\in \mathcal{U}$, $u$ sends $c_{u, v} \gets $\texttt{Enc}$(\makeAngle{ek_{u, v}}, u||v||\zeta_{u, v})$, where $\makeAngle{ek_{u, v}}\gets $\func{Key.Agr}$(\makeAngle{ek_{pri,  u}}, \makeAngle{ek_{pub, v}})$.
     
    \State $v\in \mathcal{U}$ decrypts $\zeta_{u, v}\gets$\texttt{Dec}$(\makeAngle{ek_{u, v}}, c_{u,v})$, and stores $(u, \zeta_{u, v})$.
    
    \State \textbf{Step 3 - Split Finding:}
    \State $\mathcal{S}$ randomly selects a feature sub-sample $\mathcal{Q}'$ from the full feature set $\mathcal{Q}$. 
    \State $\mathcal{S}$ invokes \protocol{SecFind}$(\mathcal{Q}', \mathcal{U})$ to determine the current optimal split. 
    \State Repeat \textbf{Step 3} until reaching the termination condition. 
  \end{algorithmic}
\end{algorithm*}

\subsection{Secure Split Finding \protocol{SecFind}}\label{sub_splitfind}
The most important operation of the CART building in XGBoost is to find the optimal split from all candidate splits to branch the leaf node. 
The candidate splits are evaluated with the split scores computed by Eq.~\ref{eq_score}.
The optimal split is the split with the maximum score.
In \sysname, split finding is implemented by \protocol{SecFind}, presented in Protocol~\ref{SecFind}.
Details of \protocol{SecFind} are as follows.
    
First, $u\in \mathcal{U}$ computes the gradients of the local training samples ($h_i$ and $w_i$, defined in Eq.~\ref{eq_score}).
Then, $\mathcal{S}$ invokes \protocol{SecAgg} twice to get the aggregation of the two kinds of gradients $H$ and $G$.
Next, for each given candidate feature $\alpha\in \mathcal{Q}'$, 
    $\mathcal{S}$ enumerates all possible candidate splits and publishes them to $\mathcal{U}$.
Similar to the above aggregation process for $H$ and $G$,
    $\mathcal{S}$ collects the left-child gradient aggregation results for each candidate split.
The aggregation results are used to compute the score for each candidate split.
When the iteration is terminated, 
    \protocol{SecFind} returns the split with the maximum score and its corresponding feature.
Intuitively, with the optimal split, $\mathcal{S}$ can add a new branch in current CART by splitting an old leaf node into two new leaf nodes.
Moreover, if the termination condition is reached after the splitting, $\mathcal{S}$ extra computes the weights of the leaf nodes with the aggregated gradients by Eq.~\ref{eq_weight}.

\begin{algorithm}[ht!]
  \caption{Secure Split Finding (\protocol{SecFind})}
  \label{SecFind}
  \begin{algorithmic}[1]
    \Require
      The sub-sampled feature set $\mathcal{Q}'$; 
      the user set $\mathcal{U}$; 
      $u\in \mathcal{U}$ holds a set of shares about other user's private mask keys $\{\zeta_{v, u}| v\in \mathcal{U}/u\}$
    
    \State $u\in \mathcal{U}$ computes $h_u\gets \sum_{i = 1}^{|D_u|} h_i$ and $w_u\gets \sum_{i = 1}^{|D_u|} w_i$.
        
    \State $\mathcal{S}$ invokes 
    $H\gets $\protocol{SecAgg}$(\mathcal{S}, \mathcal{U}, \{h_u| u\in \mathcal{U}\}, \{\zeta_{v, u}|u\in \mathcal{U}, v\in \mathcal{U}/u\})$ 
    and $W\gets $\protocol{SecAgg}$(\mathcal{S}, \mathcal{U}, \{w_u| u\in \mathcal{U}\},$ $ \{\zeta_{v, u}|u\in \mathcal{U}, v\in \mathcal{U}/u\})$.

    \For {$1 \leq q \leq \delta$}
        \State $\mathcal{S}$ enumerates every possible candidate split $A_q = \{a_1, a_2, ..., a_m\}$ for feature $\alpha_q\in \mathcal{Q}'$ and publishes them to $\mathcal{U}$. For each $a_r\in A_q$, take the following steps.
        
        \State Based on the candidate splits, $u\in \mathcal{U}$ computes $h_{u, L}\gets \sum_{i = 1}^{|D_L|} h_{i}$ and $w_{u, L}\gets \sum_{i = 1}^{|D_L|} w_{i}$.
        
        \State $\mathcal{S}$ invokes
        $H_{L}\gets $\protocol{SecAgg}$(\mathcal{S}, \mathcal{U}, \{h_{u, L}| u\in \mathcal{U}\},$ $ \{\zeta_{v, u}|u\in \mathcal{U}, v\in \mathcal{U}/u\})$ 
        and $W_{L}\gets $\protocol{SecAgg}$(\mathcal{S}, \mathcal{U},$ $ \{w_{u, L}| u\in \mathcal{U}\}, \{\zeta_{v, u}|u\in \mathcal{U}, v\in \mathcal{U}/u\})$.
        
        \State $\mathcal{S}$ computes $H_R\gets H - H_L$ and $W_R\gets W - W_L$.
            
        \State $score\gets max(score, \frac{W_L^2}{H_L + \lambda} + \frac{W_R^2}{H_R + \lambda} - \frac{W^2}{H + \lambda})$. 
    \EndFor
    
    \State \Return The optimal split with maximum $score$.
  \end{algorithmic}
\end{algorithm}

\subsection{Robustness against User Dropout.} \label{sec_UserDrop}
Two possible cases of user dropout in \sysname are discussed as follows. 

\vspace{-0.2cm}
\textbf{Case 1:} 
A user $u_0$ drops out at the first or second step of \protocol{SecBoost}. 
In such a case, the user becomes illegal.
$\mathcal{S}$ refuses $u_0$ to be involved in the current round of training and replaces the user by another active user if possible.
    
\textbf{Case 2:} A user $u_0$ drops out during the secure aggregation process of split finding.
$\mathcal{S}$ recovers the private mask key of $u_0$ and removes $u_0$ from $\mathcal{U}$,
that is, the active user list becomes $\mathcal{U}'\subseteq \mathcal{U}$ and $u_0\in (\mathcal{U}$ $\backslash$ $\mathcal{U}')$.
To recover the private mask key of $u_0$,
    $\mathcal{S}$ collects the shares of its private mask key from at least $t$ users, i.e., $\{\zeta_{u_0, v}| v\in \mathcal{U}'\}$ and $|\mathcal{U}'|>t$.
Then, $\mathcal{S}$ recovers the private mask key of $u_0$ through $\makeAngle{sk_{pri, u_0}}\gets $\texttt{SS.Recon}$(\{\zeta_{u_0, v}| v\in \mathcal{U}'\}, t)$. 
Using $\makeAngle{sk_{pri, u_0}}$, $\mathcal{S}$ computes the shared mask keys that $u_0$ uses to mask the gradients, $\{\makeAngle{sk_{u_0, v}}| v\in \mathcal{U}\}$, where $\makeAngle{sk_{u_0, v}}\gets $\texttt{KEY.Agr}$(\makeAngle{sk_{pri, u_0}}, \makeAngle{sk_{pub, v}})$.
Finally, $\mathcal{S}$ adds the shared mask keys to the aggregated result, shown in line 8, Protocol~\ref{pro_secagg}.
In this way, $\mathcal{S}$ can still get the correct aggregation result of remaining active users' gradients, whose correctness has been discussed in Section~\ref{sec_technical}.

\section{Security Analysis}
\label{sec_analysis}
In this section, we discuss the security of \sysname for secure aggregation and XGBoost training.

\subsection{Security of \protocol{SecAgg}}
For \protocol{SecAgg}, we first present how it achieves our security goals, and then, give a more formal security proof in the next sub-section.

\protocol{SecAgg} achieves forced aggregation because from the analysis of correctness, ignoring any user's data makes $\mathcal{S}$ unable to get a meaningful result.
Then, consider a single user's masked value $[\![x_u]\!]$ derived by Eq.~\ref{eq_secmask}.
For an eavesdropper or a malicious user, $[\![x_u]\!]$ is a ciphertext encrypted with the server's public key, which is semantically secure based on the security of Bresson's cryptosystem~\cite{bresson2003simple}.
If a malicious central server is included, there are two conditions required to be discussed:
a) when the user does not drop out, the adversary can decrypt $[\![x_u]\!]$ but cannot obtain $x_u$ that is masked by $\Upsilon_u$.
b) when the user drops out, the adversary can access both $\Upsilon_u$ and $\makeAngle{sk_{pri, \mathcal{S}}}$.
However, since $g^{r_u}$ is unknown, $[\![x_u]\!]$ is still undecipherable.
Further, assume that the central server is a more active attacker (out of the scope of our security model).
In such a case, the central server can cheat the user to upload $g^{r_u}$ by sending a forged active user list.
We can defend the attack by letting the user sign the active user list $\mathcal{U}'$ and exchanging it with other users.
Thus, by checking the consistency of $\mathcal{U}'$, the user can defend the active attack.

\subsection{Security of \sysname}
The security of \sysname for XGBoost training is determined by three protocols, \protocol{SecBoost}, \protocol{SecFind} and \protocol{SecAgg}.
To prove the protocols' security, we adopt the standard formal definition of security \textit{Definition 2} \cite{ma2019lightweight}.

\noindent \textbf{Definition 2.} \textit{We say that a protocol $\pi$ is secure if there exists a probabilistic polynomial-time (PPT) simulator $\xi$ that can generate a view for the adversary $\mathcal{A}$ in the real world and the view is computationally indistinguishable from its real view.}

Moreover, our security still needs the following lemma.

\noindent \textbf{Lemma 1 \cite{bogdanov2008sharemind}.} A protocol is perfectly simulatable if all its sub-protocols are perfectly simulatable.

Interested readers can refer to \cite{bogdanov2008sharemind} for the detailed proof of \textit{Lemma 1}.
According to \textit{Lemma 1} and \textit{Definition 2}, to prove the security of \sysname, we just have to prove that all of its protocols are simulatable for a PPT simulator.
Since \protocol{SecAgg} is a sub-protocol that is frequently invoked by \protocol{SecFind}, we merge the proof of \protocol{SecAgg} into \protocol{SecFind}.
The security proofs of \protocol{SecFind} and \protocol{SecBoost} are given below. 

\noindent \textbf{Theorem 1.} \textit{For \protocol{SecFind}, there exists a PPT simulator $\xi$ that can simulate an ideal view which is computationally indistinguishable from the real view of $\mathcal{A}$.}

\noindent \textit{Proof.}
Denote the views of the user $u\in\mathcal{U}$ as $\mathcal{V}_u = $ $\{view_{u_1}, ...,view_{u_{n}}\}$.
From \protocol{SecFind}, we can derive that
$view_{u_i} = $ $\{h_{u_i},w_{u_i}, r_{u_i}, g^{r_{u_i}}, [\![h_{u_i}]\!], [\![w_{u_i}]\!], \makeAngle{sk_{u, v}},$ $ \makeAngle{sk_{pri, u_i}}, \makeAngle{sk_{pub, v}}, \makeAngle{sk_{pub, \mathcal{S}}}\}$
and
$view_{\mathcal{S}} = \{H, W, H_j,$ $ W_j, A, H_L, H_R, W_L, W_R, score, \mathcal{Y}, \mathcal{R}, \mathcal{K}, \makeAngle{sk_{pri, \mathcal{S}}}\}$, 
where $u\in \mathcal{U}$, $v\in \mathcal{U}$, $\mathcal{R} = \{[\![h_{u_i}]\!],$ $ [\![w_{u_i}]\!]| u\in \mathcal{U}'\}$, $\mathcal{R} = \{g^{r_{u_i}}| u\in \mathcal{U}'\}$, $\mathcal{K} = \{\makeAngle{sk_{pri, u_i}}| u\in \mathcal{U}/\mathcal{U}'\}$. 
$[\![h_{u_i}]\!], [\![w_{u_i}]\!]$, $\mathcal{Y}$, $\mathcal{R}$ and $\mathcal{K}$ are the variables used in \protocol{SecAgg}.

We prove the security of \protocol{SecFind} according to the universal composition theorem (UC) \cite{mohassel2017secureml}.
Assume that there is an ideal functionality $\mathcal{F}$ that can be called by a simulator $\xi$.
$\mathcal{F}$ has the ability to ideally generate uniformly random values and operate the cryptographic functions of \sysname.
We say that with $\mathcal{F}$, there exists such a simulator that can simulate both the honest entity and the corrupted (curious-but-honest) entity in \protocol{SecFind} as follows.
For the corrupted entity, $\xi$ can access all its local data, including the private keys, training samples and etc., based on our security model.
Therefore, $\xi$ can simply use the corrupted data to simulate the corrupted entity.
For the honest entity, the simulation is a little complicated.
To simulate an honest user, $\xi$ first asks $\mathcal{F}$ to generates random values as dummy function inputs $h_{u_i}$, $w_{u_i}$, $r_{u_i}$ and $\makeAngle{sk_{pri, \mathcal{S}}}$.
Then, use the dummy inputs to derive other variables to ask $\mathcal{F}$ to complete the protocol steps.
Similarly, $\xi$ can use the same way to simulate an honest server.
It is observed that the elements in $View_{u_i}$ or $View_{\mathcal{S}}$ are either the ciphertext in the Bresson's cryptosystem or random values ($h_{u_i}$ and $w_{u_i}$ are private, which can be seen as random values).
Since the Bresson's cryptosystem is semantically secure~\cite{bresson2003simple}, the dummy function outputs are computationally indistinguishable from the real ones.
Consequently, there exists a simulator that can generate the simulated views $Sim_{u_i}$ and $Sim_{\mathcal{S}}$ that are indistinguishable from $View_{u_i}$ and $View_{\mathcal{S}}$. 
Based on \textit{Definition 2}, \textit{Theorem 1} holds and \protocol{SecFind} is secure.
\QEDB

\noindent \textbf{Theorem 2.} \textit{For \protocol{SecBoost}, there exists a PPT simulator $\xi$ that can simulate an ideal view which is computationally indistinguishable from the real view of $\mathcal{A}$.}

\noindent \textit{Proof.}\quad Denote the views of the user and the central server for \protocol{SecBoost} as $\mathcal{V}_u = \{view_{u_1}, ..., view_{u_{n}}\}$ and $\mathcal{V}_{\mathcal{S}}$. 
The view of the user is
$view_{u_i} = \{\makeAngle{sk_{pri, u_i}}, \makeAngle{sk_{pub, u_i}}, \makeAngle{ek_{pri, u_i}}, \makeAngle{ek_{pub, u_i}}, c_{v, u}, \{\zeta_{v, u}|v\in \mathcal{U}'\}, view_{u_i}'\}$. 
For the central server, we have $view_{\mathcal{S}} = \{\makeAngle{sk_{pub, \mathcal{S}}}, \makeAngle{sk_{pri, \mathcal{S}}} view_{\mathcal{S}}'\}$, 
 $view_{u_i}'$ and $view_{\mathcal{S}}'$ are the views generated by \protocol{SecFind}. 
Except for the encryption keys randomly selected from $\mathbb{Z}^*_N$, the remaining elements of $view_{u_i}$ and $view_{\mathcal{S}}$ are random shares or ciphertexts encrypted with the shared key encryption algorithm. 
According to Shamir's secret sharing theory~\cite{mceliece1981sharing}, the shares can be regarded as random values uniformly selected from $\mathbb{Z}^*_p$.
To simulate the ciphertexts and random shares, a simulator $\xi$ can use the ideal functionality $\mathcal{F}$ defined in the proof of \protocol{SecFind} to generate random values as dumpy cryptographic function's inputs.
Since the shared key encryption algorithm is assumed to be indistinguishable under a chosen plaintext attack, the corresponding dumpy outputs cannot be computationally distinguished.
Moreover, $view_{u_i}'$ and $view_{\mathcal{S}}'$ have been proved to be simulatable in the proof of \textit{Theorem 2}.
Thus, there exists a simulator $\xi$ that can simulate a corrupted entity or an honest entity of \protocol{SecBoost} and the simulated view is computationally indistinguishable from the real view.
Based on \textit{Definition 2}, \textit{Theorem 2} holds and \protocol{SecBoost} is secure. \QEDB

According to \textit{Lemma 1} and the above proofs, it is concluded that \sysname is a simulatable system. 
Based on the formal definition of security given in \textit{Definition 2}, \sysname is secure.

\section{Performance Evaluation}
\label{sec_experiments}
In this section, 
    we conduct extensive experiments to evaluate the effectiveness and efficiency of \sysname.

\subsection{Experiment Configuration}
To evaluate \sysname, we ran single-threaded simulations on a Windows desktop with an Intel Core i7-8565U CPU @1.8Ghz and 16G RAM.
The programs are implemented in Python and C++. 
Two standard datasets are used in the experiments,
    ADULT\footnote{ADULT: https://www.csie.ntu.edu.tw/~cjlin/libsvmtools/datasets} and MNIST\footnote{MNIST: http://yann.lecun.com/exdb/mnist/}, 
    both of which are commonly used to evaluate the performance of federated learning schemes~\cite{mcmahan2016communication, bonawitz2017practical, mcmahan2017learning}.
The Bresson's cryptosystem is conducted with a key size of 512 bits.

The shared-key encryption is operated by 128-bit {AES-GCM}~\cite{bellare2016multi}.
Given each dataset,
    the instances are averagely and randomly assigned to each user with no overlap. 
User dropout is assumed to occur every 10 rounds of training in our experiment.
That is,
    0\%, 10\%, 20\%, 30\% of users are randomly selected to be disconnected at each $10^{th}$ round of training.

\subsection{Evaluation of \sysname}
To assess the performance of \sysname, we first evaluate its effectiveness with ADULT and MNIST.
Then, we experiment with its runtime to find an optimal split to test its efficiency.

\begin{figure}[htbp]
\centering
\subfigure[Accuracy with different user dropout rates for ADULT.]{
\begin{minipage}[t]{0.45\linewidth}
\centering
\includegraphics[scale=0.175]{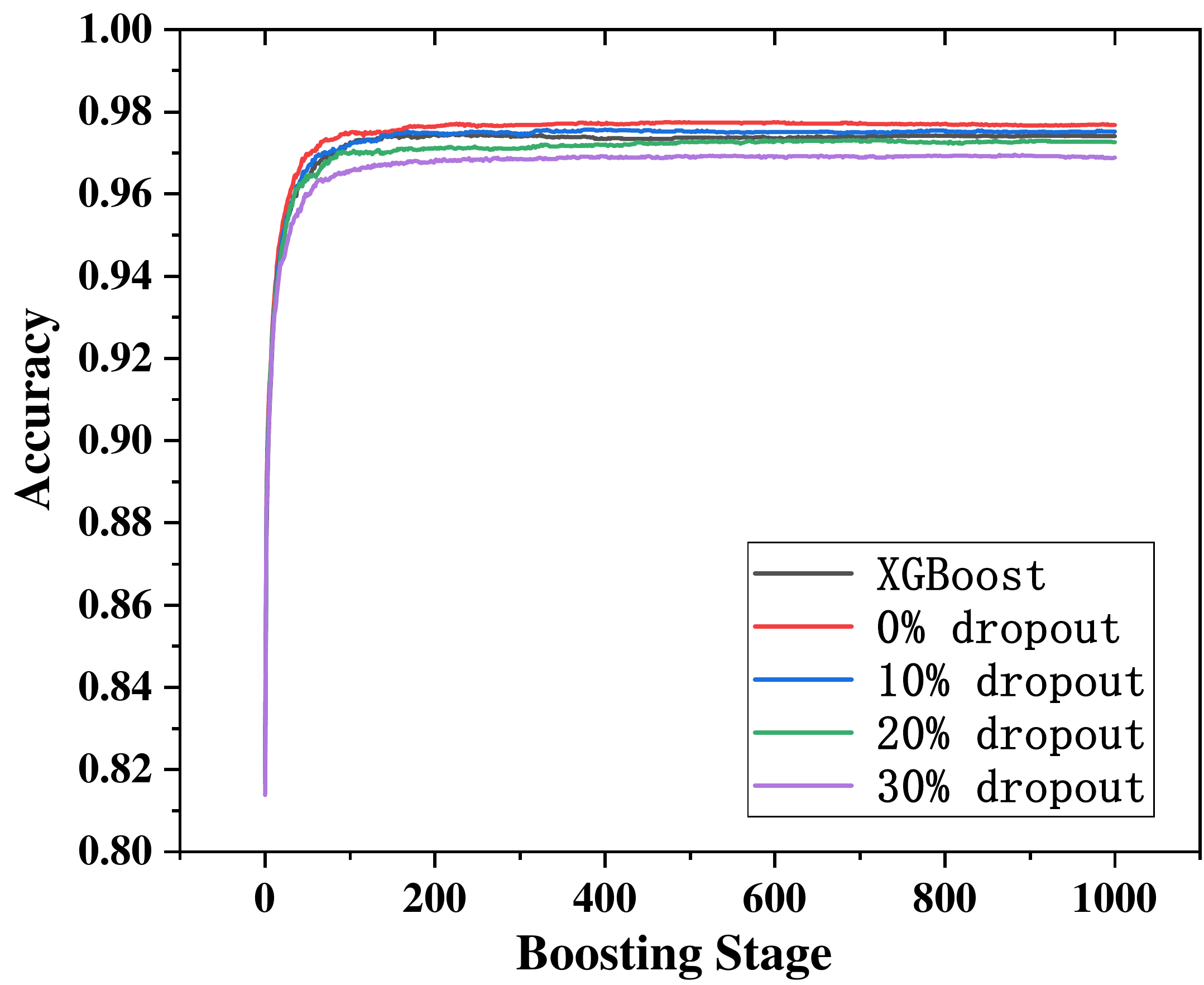}
\label{Accuracy_Adult}
\end{minipage}
}%
\hfill
\subfigure[Loss with different user dropout rates for ADULT.]{
\begin{minipage}[t]{0.45\linewidth}
\centering
\includegraphics[scale=0.18]{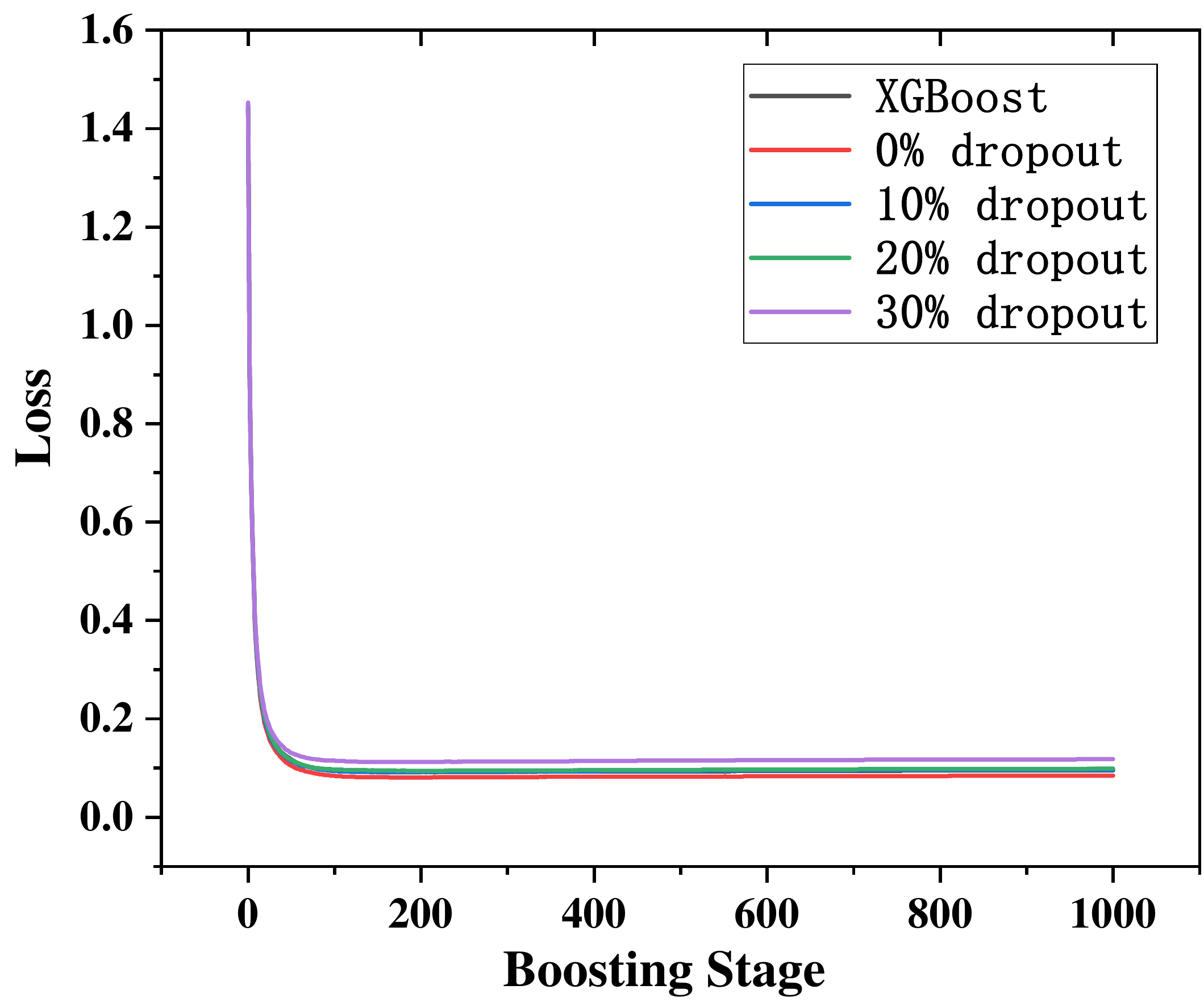}
\label{Loss_Adult}
\end{minipage}
}%
\hfill
\subfigure[Accuracy with different user dropout rates for MNIST.]{
\begin{minipage}[t]{0.45\linewidth}
\centering
\includegraphics[scale=0.175]{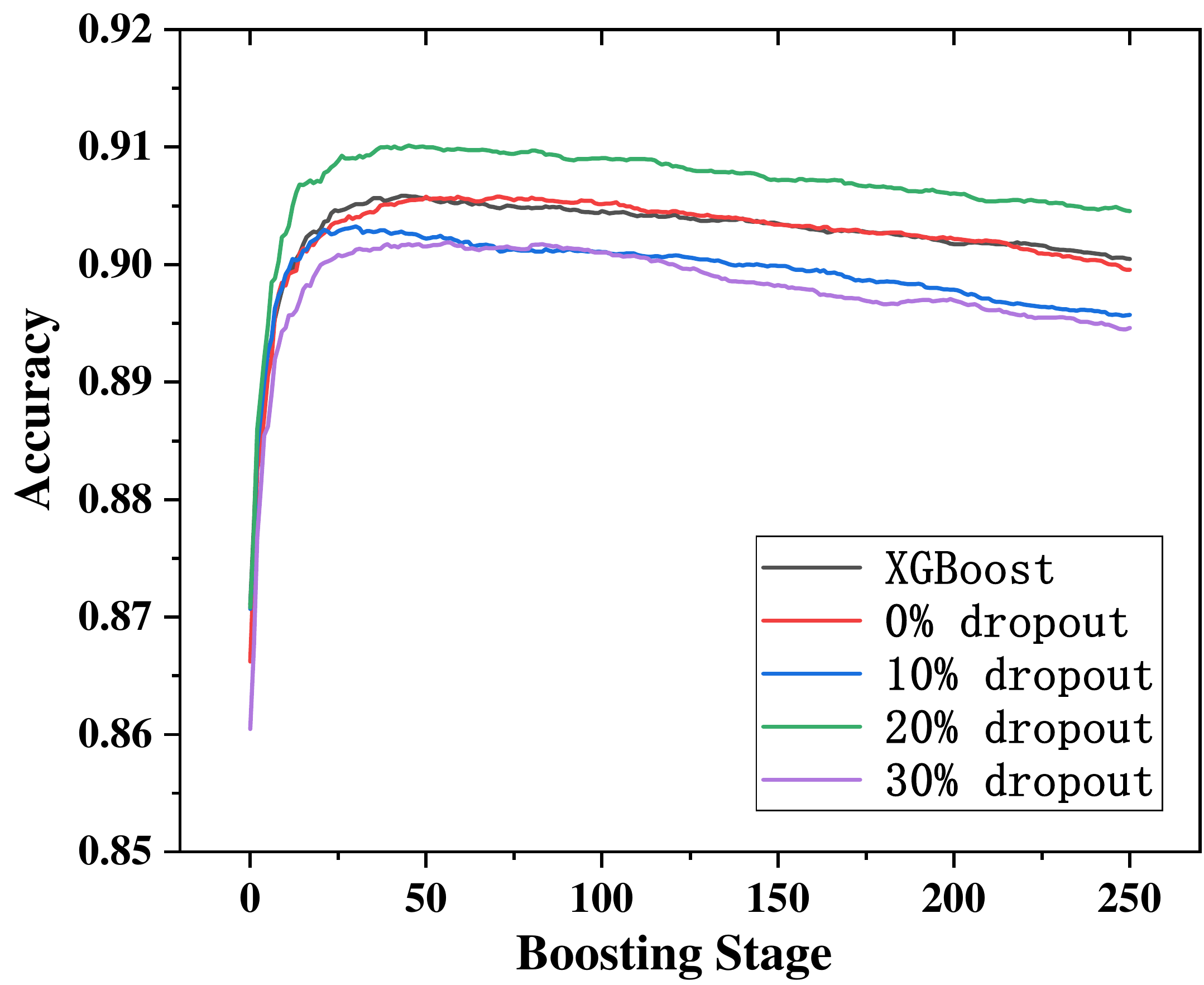}
\label{Accuracy_MNIST}
\end{minipage}
}%
\hfill
\subfigure[Loss with different user dropout rates for MNIST.]{
\begin{minipage}[t]{0.45\linewidth}
\centering
\includegraphics[scale=0.175]{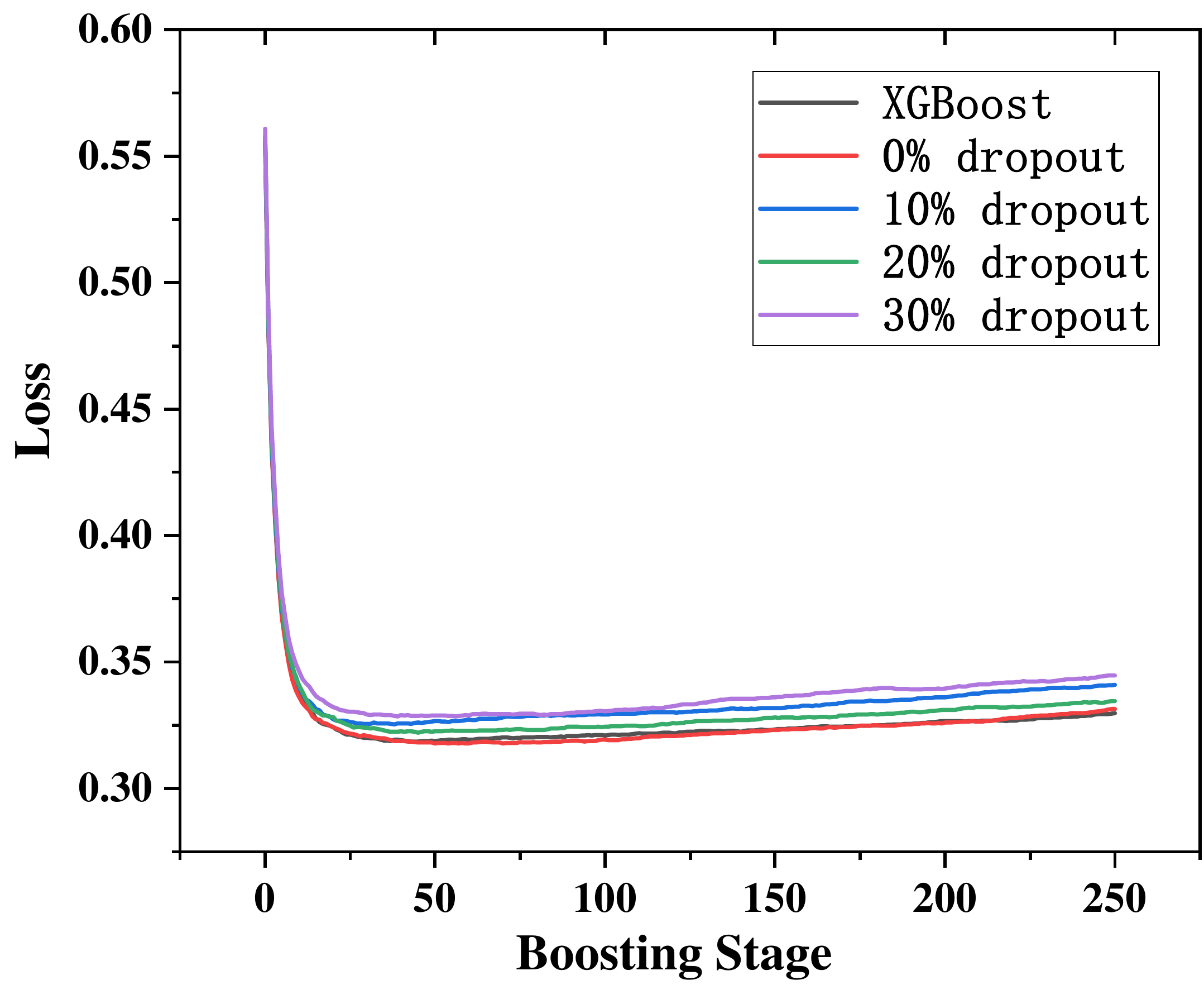}
\label{Loss_MNIST}
\end{minipage}
}%
\caption{Accuracy and loss for each round of training with MINST and ADULT. Different lines show different dropout rates.}
\label{Auc_Loss}
\end{figure}

The effectiveness of \sysname is assessed with two indicators that are commonly used to evaluate a machine learning model, namely classification accuracy and loss. 
Fig.~\ref{Auc_Loss} presents the accuracy and loss for each round of training in \sysname.
More specific,
    Fig.~\ref{Accuracy_Adult} and Fig.~\ref{Loss_Adult} describe the accuracy and loss of ADULT, 
    and Fig.~\ref{Accuracy_MNIST} and Fig.~\ref{Loss_MNIST} show the result of MNIST.
For ADULT, the accuracy peaks after around 100 rounds. 
For MINST, the convergence speed is faster, peaking at around the 20$^{th}$ rounds.
Compared with the non-federated XGBoost,
    \sysname only introduces the accuracy loss with less than 1\%.
Consider the user dropout rate increased from 0\% to 30\%,
    \sysname is robust against the user changes.
The performance decrease is mainly due to the loss of data caused by the user dropout.
In Table~\ref{table_overhead}, 
    we list the runtime and communication cost of different stages to execute \protocol{SecBoost} to find an optimal split in \sysname without user dropout.
The processed data in the experiment is ADULT.
The user number is set to 500. 
The system setup cost is ignored.
The result indicates that the main overhead in \sysname is caused by the split finding step, 
    because numerous secure aggregation protocols are invoked.
Therefore, in the next section, we comprehensively analyze the efficiency of \protocol{SecAgg}.

\vspace{-0.2cm}
\renewcommand\arraystretch{1.5}
\begin{table}[!htbp]
\centering
\caption{\protocol{SecBoost} runtime in different Stages without User Dropout}
\begin{tabular}{c|c|c|c|c|c}

\hline
\multicolumn{1}{c|}{\multirow{2}{*}{Stage}}     & \multicolumn{2}{c|}{RunTime (s)} & \multicolumn{2}{c}{Communication (MB)}\\

\cline{2-5}
& $\mathcal{U}$     &    $\mathcal{S}$  & $\mathcal{U}$     &    \multicolumn{1}{c}{$\mathcal{S}$} \\



\cline{1-5}
Mask Key Sharing    & 0.89      & N.A.         & 0.06  & \multicolumn{1}{c}{N.A.}\\

\hline
Split Finding       & 1.22      & 249.97       & 26.86 & \multicolumn{1}{c}{46.25}\\

\hline
Total Cost          & 2.11      & 249.97      & 0.05    & \multicolumn{1}{c}{13.57} \\

\hline
\end{tabular}
\label{table_overhead}
\end{table}


\subsection{Efficiency Analysis of \protocol{SecAgg}}
To further assess the efficiency of \sysname, we simulate the runtime and communication costs of \protocol{SecAgg} under different numbers of users, input sizes and dropout rates.

\vspace{0.1cm}
\noindent
\textbf{Theoretical Analysis.}
Suppose the transmitted data is a vector with $m$ entries.
The length of $N$ is $\mathcal{N} = \lfloor \log_2 N \rfloor$.
The user number is $n$. 
For communication, each user sends one random seed, $m$ ciphertexts and $\mathcal{O}(n)$ shares for private mask key and dropout users, whose complexity is $\mathcal{O}(m + n)$.
The server receives the user's ciphertexts, the random seeds and key shares of dropout users, whose overhead is $\mathcal{O}(nm + n^2)$.
For computation, suppose the modular exponentiation costs $1.5\mathcal{N}$ multiplications~\cite{knuth2014art}.
Each user computes $2m$ times modular exponentiation and $\mathcal{O}(n^2)$ multiplications for sharing the private mask key, which takes $\mathcal{O}(m\mathcal{N} + n^2)$ time.
The server conducts $mn$ times modular exponentiation to decryption and $\mathcal{O}(n^2)$ multiplications for data recovering of the dropout user, which takes $\mathcal{O}(mn\mathcal{N} + n^2)$ time.

\begin{figure}[htbp]
\centering
\subfigure[Runtime as the user number increases.]{
\begin{minipage}[t]{0.45\linewidth}
\centering
\includegraphics[scale=0.175]{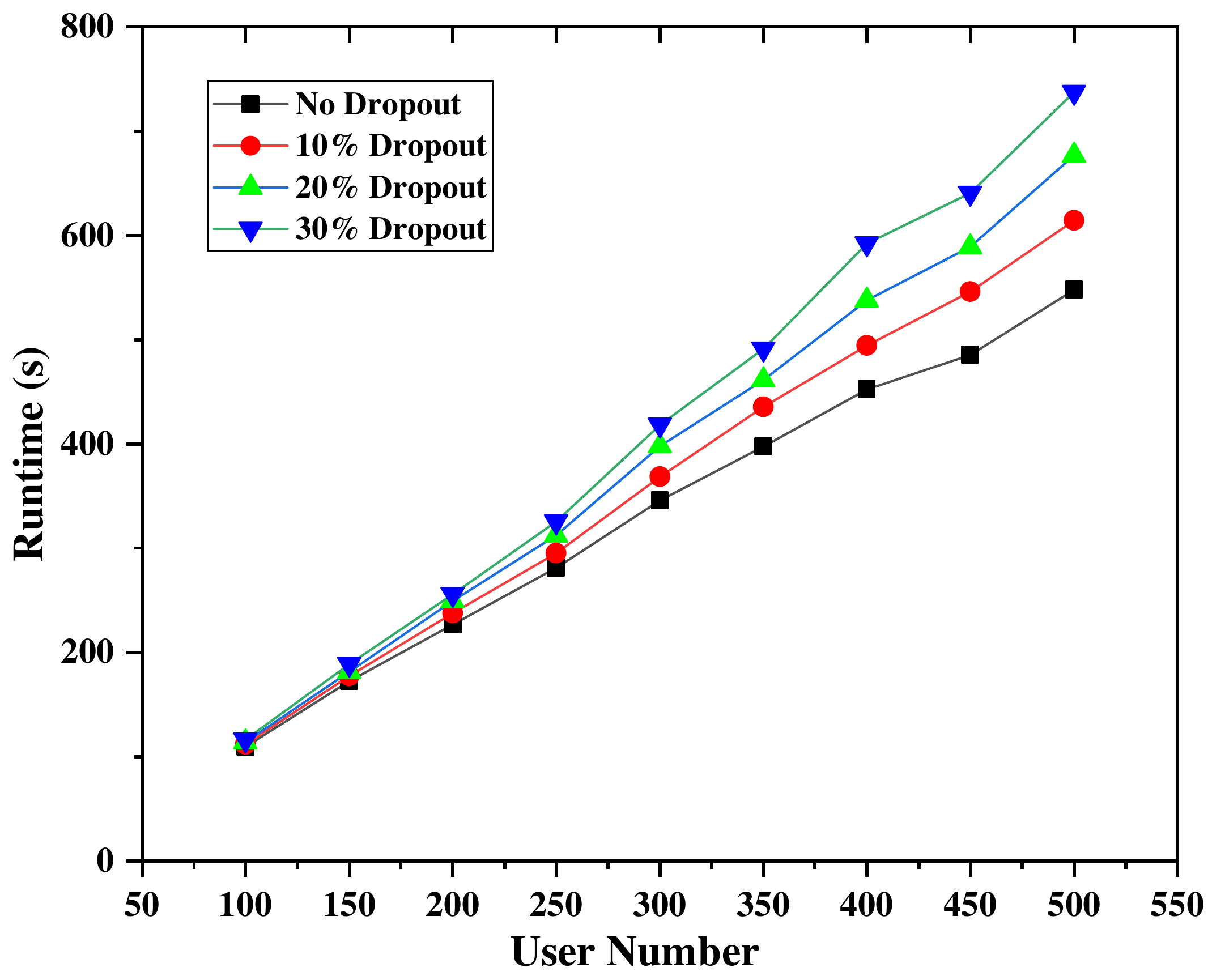}
\label{fig_server_runtime_user}
\end{minipage}
}%
\hfill
\subfigure[Communication overhead as the user number increases]{
\begin{minipage}[t]{0.45\linewidth}
\centering
\includegraphics[scale=0.18]{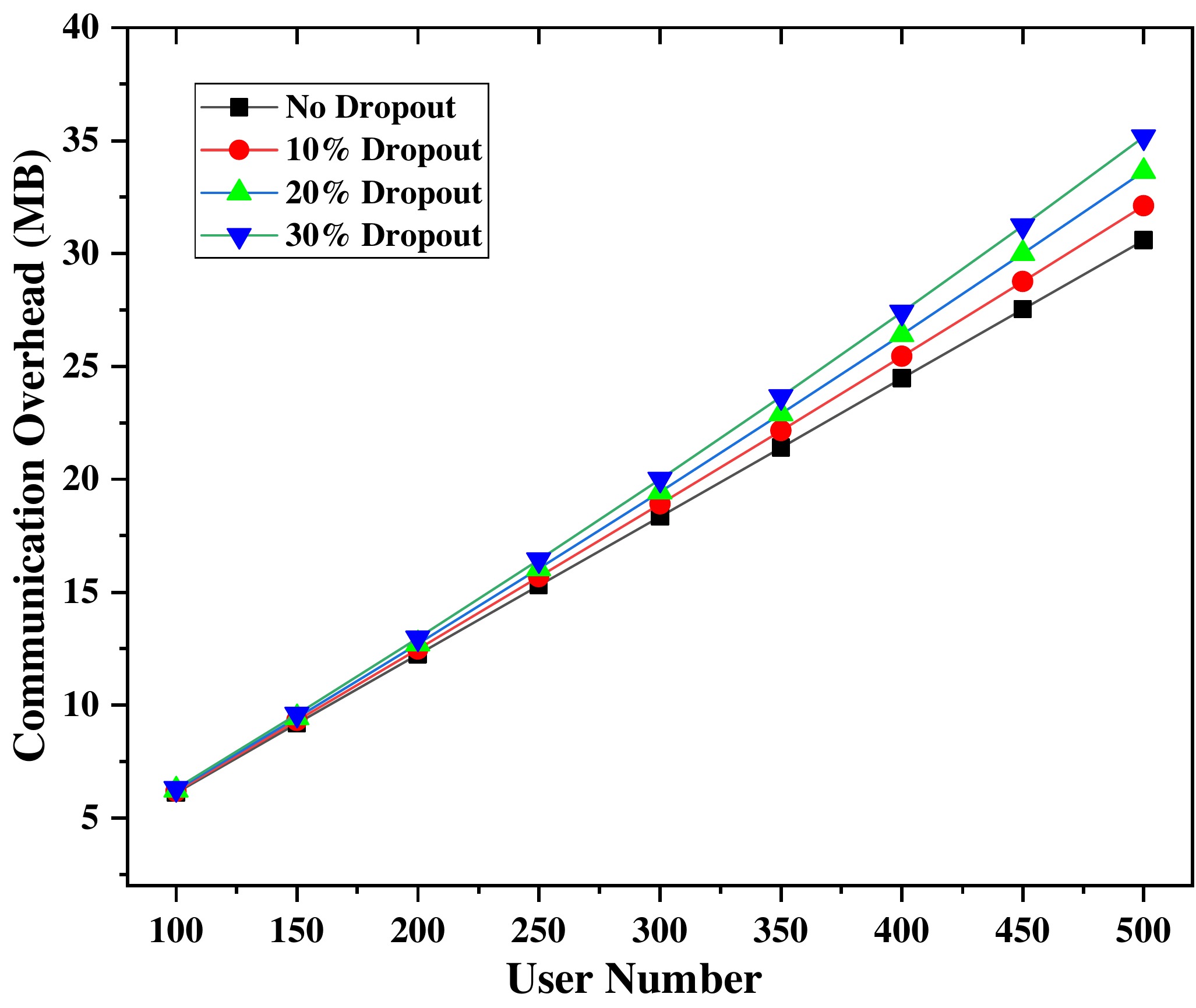}
\label{fig_server_comm_user}
\end{minipage}
}%
\caption{Efficiency evaluation of the central server with fixed input size 500. Different lines show different dropout rates.}
\label{fig_server_efficiency_user}
\end{figure}

\begin{figure}[htbp]
\centering
\subfigure[Runtime as the input size increases.]{
\begin{minipage}[t]{0.45\linewidth}
\centering
\includegraphics[scale=0.175]{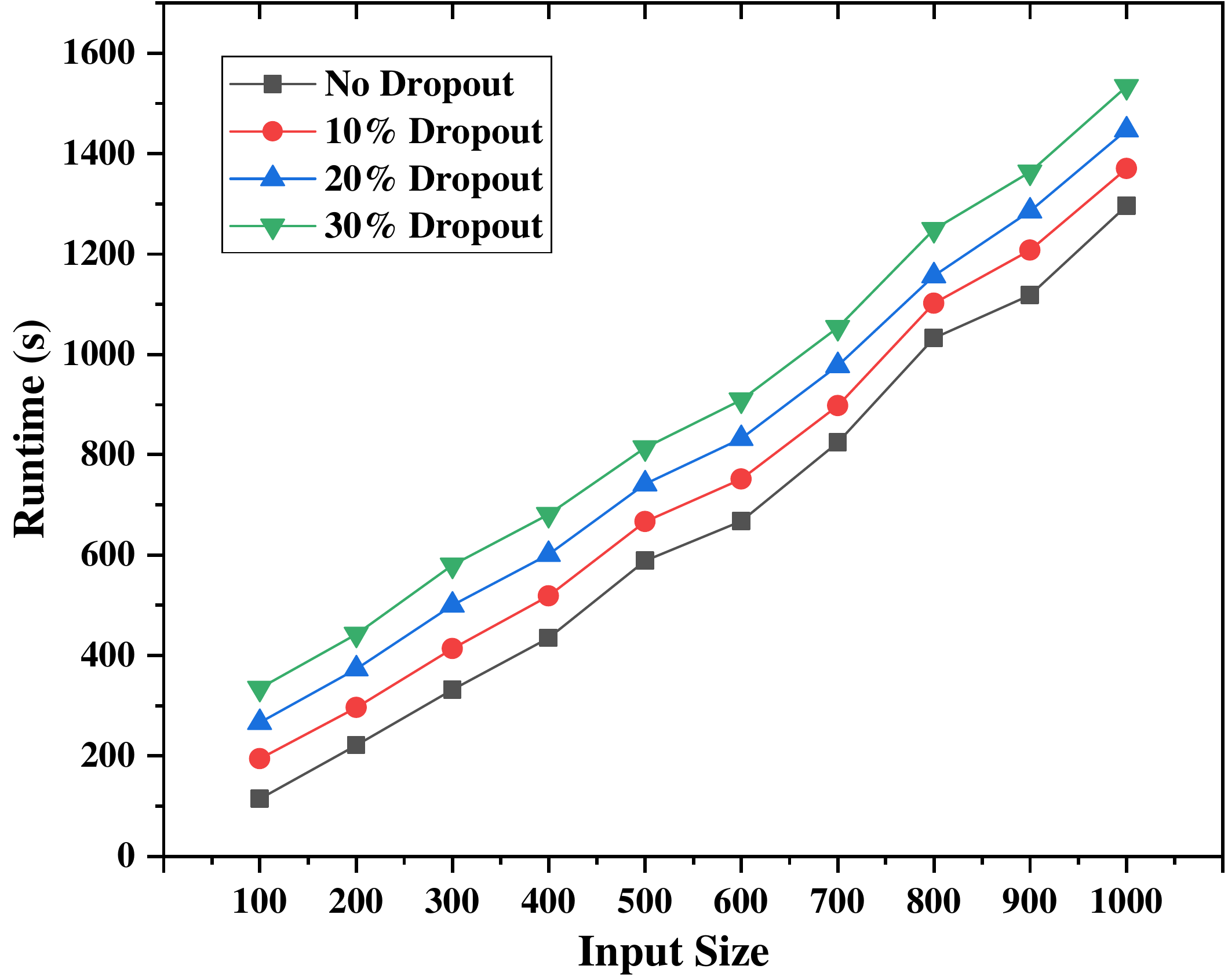}
\label{fig_server_runtime_input}
\end{minipage}
}%
\hfill
\subfigure[Communication overhead as the input size increases.]{
\begin{minipage}[t]{0.45\linewidth}
\centering
\includegraphics[scale=0.18]{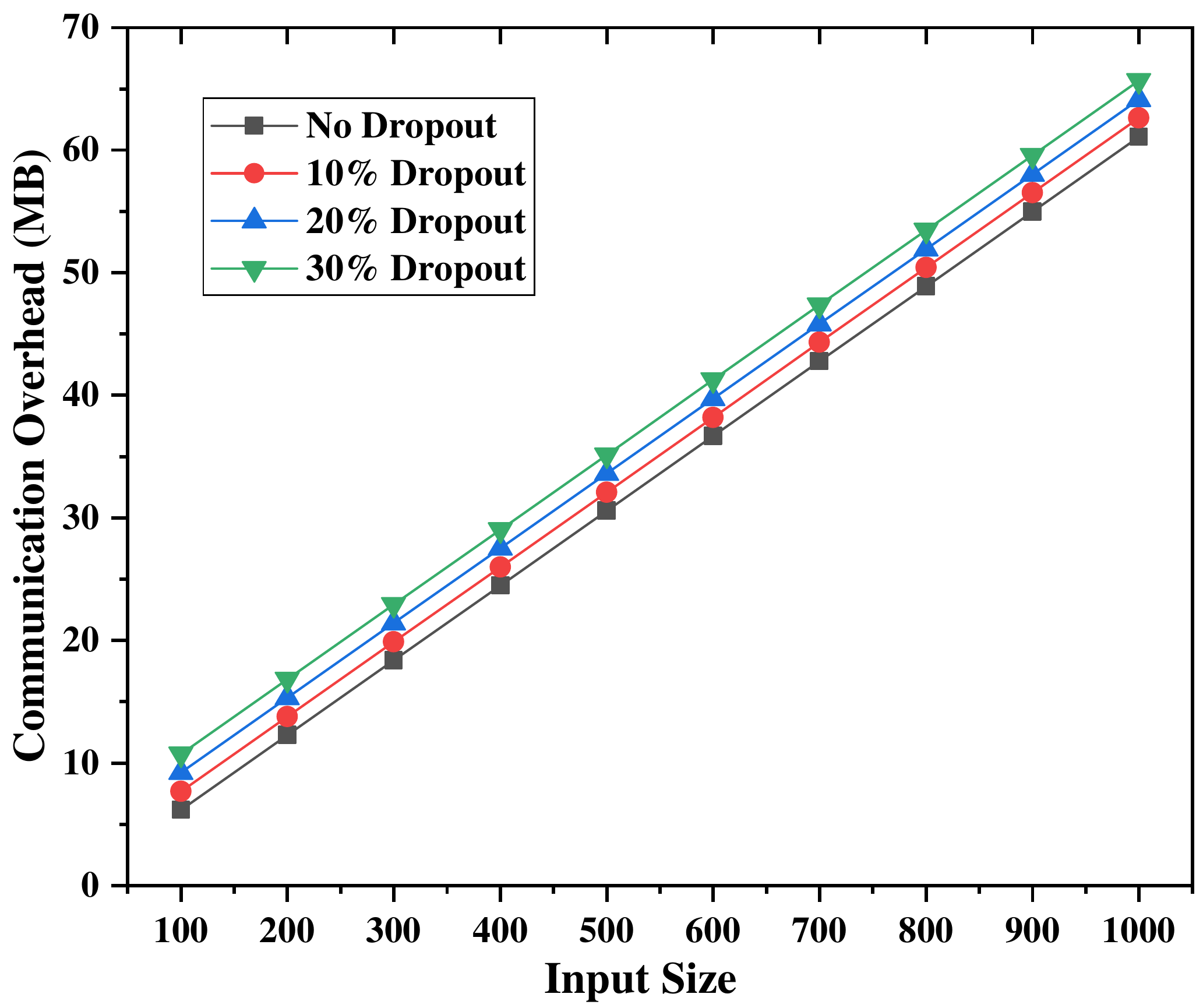}
\label{fig_server_comm_input}
\end{minipage}
}%
\caption{Efficiency evaluation of the central server with fixed user number 500. Different lines show different dropout rates.}
\label{fig_server_efficiency_input}
\end{figure}

\begin{figure}[htbp]
\centering
\subfigure[Runtime per user as the user number increases.]{
\begin{minipage}[t]{0.45\linewidth}
\centering
\includegraphics[scale=0.175]{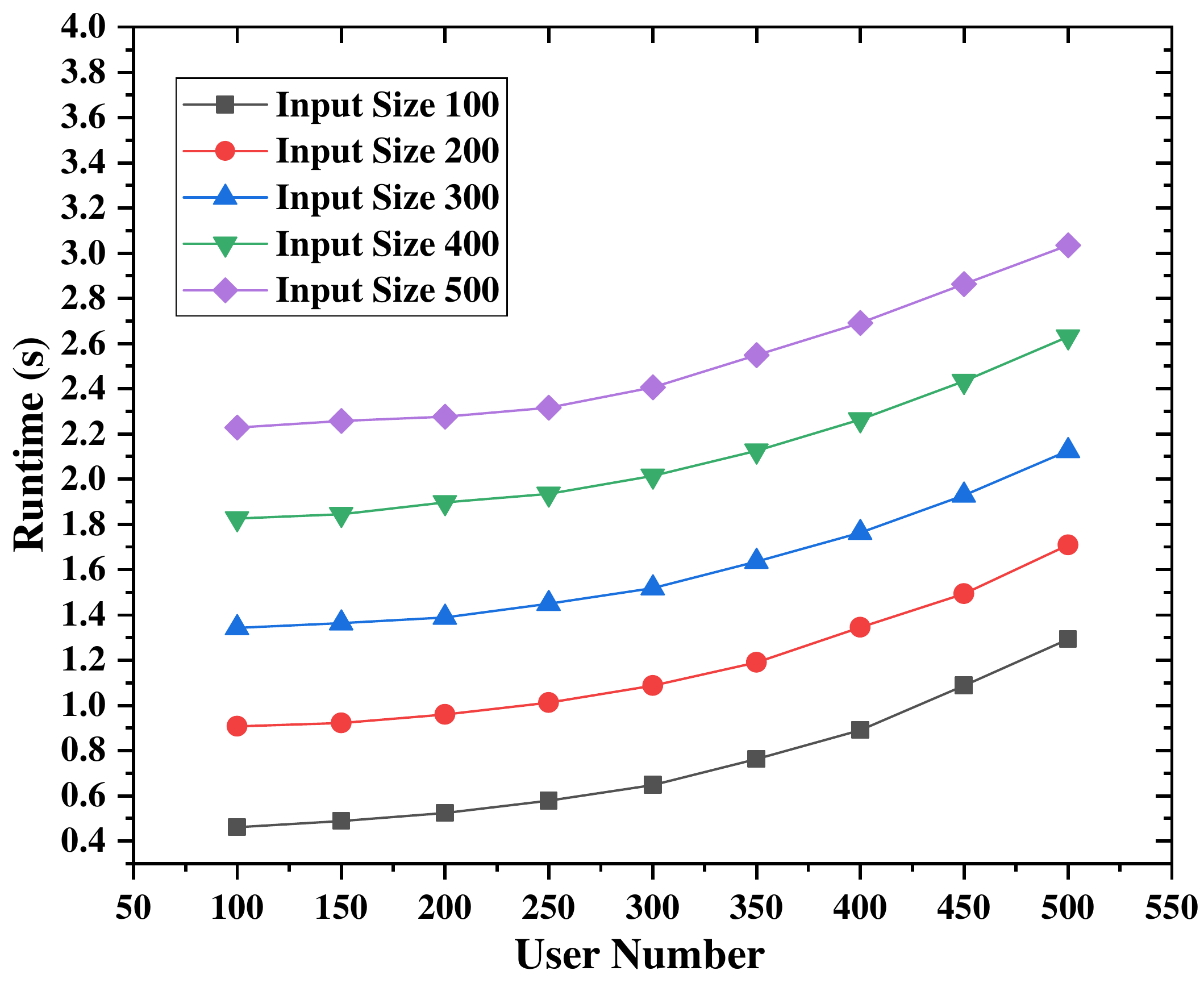}
\label{fig_user_runtime}
\end{minipage}
}%
\hfill
\subfigure[Communication overhead per user as the user number increases.]{
\begin{minipage}[t]{0.45\linewidth}
\centering
\includegraphics[scale=0.18]{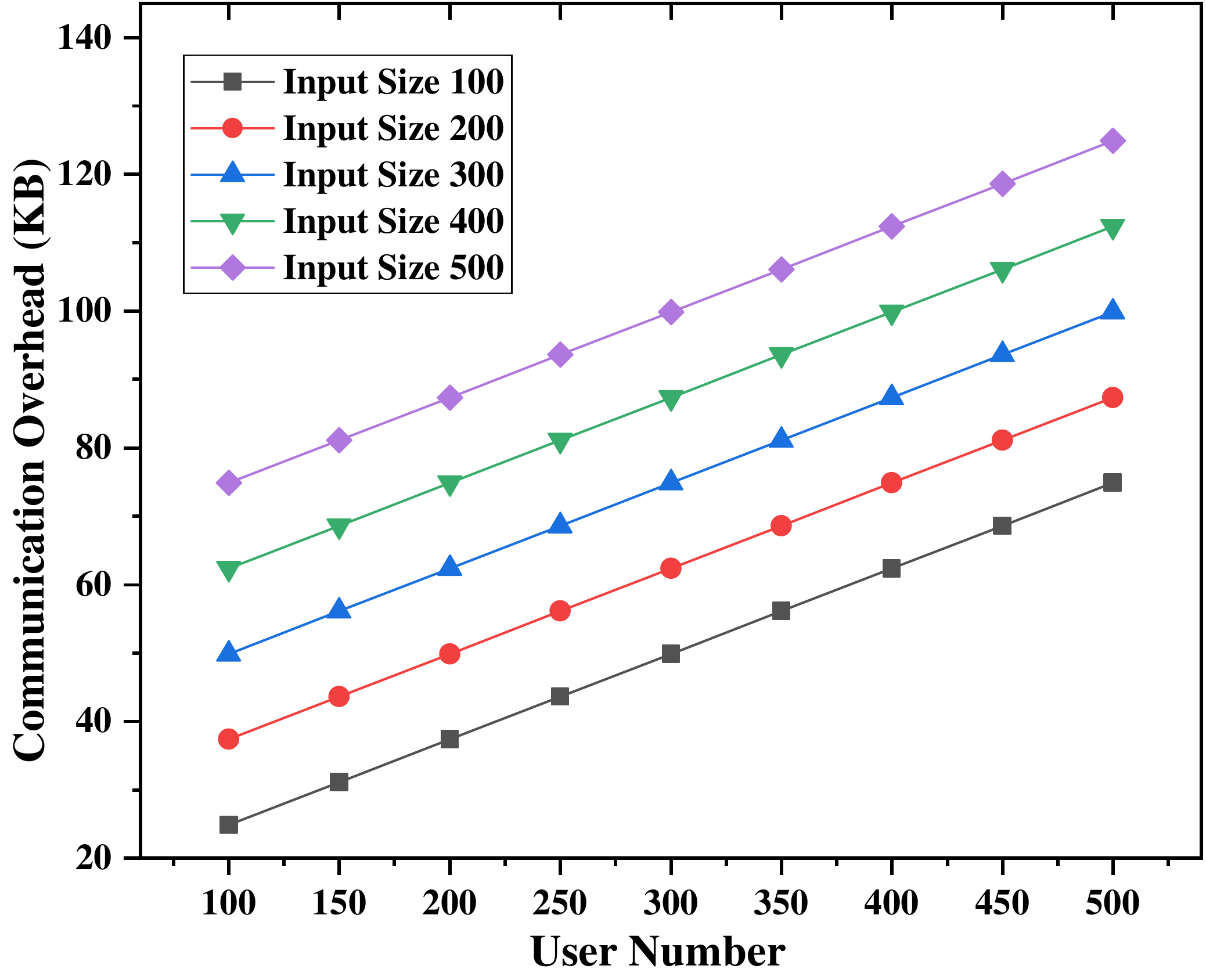}
\label{fig_user_comm}
\end{minipage}
}%
\caption{Efficiency evaluation of the user without user dropout.}
\label{fig_user_efficiency}
\end{figure}

\vspace{0.1 cm}
\noindent
\textbf{Impact of Users.}
The runtime and communication overhead of the central server and the user with different user numbers are plotted in Fig.~\ref{fig_server_efficiency_user}, Fig.~\ref{fig_user_efficiency}, respectively.
The input size is fixed to 500.
The steps of setup and the shared mask keys generation can be previously completed, therefore, they are not included in the evaluations.
We omit the plot of the runtime and communication overhead of the user with different dropout rates, as the user just has to send the private mask key shares of the dropout users, which has little impact on the metrics.

As shown in Fig.~\ref{fig_server_runtime_user} and Fig.~\ref{fig_user_runtime}, it is shown that the runtime for the central server and the user grows with the increasing user number.
For the central server, the runtime is mostly spent on the modular exponentiation to decrypt the masked aggregation result.
For the user, most computational costs are also focused on the modular exponentiation but for masking the gradient.
The user dropout rate has a significant influence on the runtime of the central server because the operation for recovering the shared mask key of the dropout users involves the costly modular exponentiation.
Fig.~\ref{fig_server_comm_user} and Fig.~\ref{fig_user_comm} illustrate the communication overhead for the user and the central server.
The communication overhead for both the user and the central server also linearly increases as the user number increases.
As the dropout rate grows, the communication overhead of the central server is barely influenced because only a little overhead increment is caused by collecting the private mask key shares for the dropout user.

\vspace{0.1cm}
\noindent
\textbf{Impact of Input Size.}
The runtime and communication overhead of the user and the central server with different input sizes are plotted in Fig.~\ref{fig_server_efficiency_input} and Fig.~\ref{fig_user_efficiency}, respectively.
In XGBoost, the input size is equal to the multiplication of the sub-sampled feature number and the enumerated candidate split number.
In the experiments, the number of users is fixed to $500$.
When the input size increases from 100 to 1000,
    the runtime cost for each user increases because of the masking operation, illustrated in Fig.~\ref{fig_user_runtime}.
The growth of the central server's runtime is mainly caused by the more masked aggregation result required to be decrypted.
As a larger scale of inputs is involved, 
    the communication overhead of the central server is expanded, shown in Fig.~\ref{fig_server_comm_input}.
For the user,
    Fig.~\ref{fig_user_comm} shows that the communication overhead is linearly influenced by the input size.
Compared with the number of users, the input size has a less obvious influence on the runtime and communication overhead, which is consistent to our theoretical analysis result given in Section~\ref{sec_technical}.
\renewcommand\arraystretch{1.5}
\begin{table}[!htbp]
\centering
\caption{Efficiency Comparison with fixed user number 500, input size 500 and no user dropout}
\begin{tabular}{c|c|c|c|c}

\hline
\multicolumn{2}{c|}{}     & \sysname & \othersys{IBMHom}~\cite{truex2018hybrid} & \othersys{FPE}~\cite{abdalla2018multi}\\

\hline
\multirow{2}{*}{Rt. }    &  Server & 557.97 & 596.03 & 952.07\\

\cline{2-5}
& User  &  3.03  & 4.67  & 2.55 \\

\hline
\multirow{2}{*}{Comm.}    &  Server & 30.57 & 67.13 & 30.58\\

\cline{2-5}
& User  &  0.12  & 0.18  & 0.08   \\

\hline
\multicolumn{5}{c}{Rt. $\to $ Runtime (s); Comm.$\to$ Communication overhead (MB)}\\

\end{tabular}
\label{table_efficiency_com}
\end{table}

\textbf{Comparison.}
Besides \othersys{IBMHom}, we compare the efficiency of \sysname with the functional Paillier encryption based secure aggregation scheme (abbreviated as \othersys{FPE})~\cite{abdalla2018multi}.
We implement all the three methods in our desktop, referring to the python library for Paillier's cryptosystem\footnote{https://github.com/data61/python-paillier}.
Moreover, for fairness, we cancel the noise addition operation of \othersys{IBMHom}.
Considering the secure aggregation, the cancellation does not influence the security of \othersys{IBMHom}.
The threshold of \othersys{IBMHom} is set to $0.6 \times n$. 
Table~\ref{table_efficiency_com} summarizes the comparison result.
It is observed that compared with \othersys{IBMHom}, \sysname outperforms it in all indicators.
Note that during the experiments, we find that the server in \othersys{IBMHom} requires very little time to decrypt the aggregation result when the user number is small, e.g., the user number $n$ is less than 100.
This is because, for the HE algorithm used in \othersys{IBMHom}, the complexity of the exponent length involved in the decryption is $\mathcal{O}(n \log n)$, not fixed $\mathcal{O}(\log N)$ in \sysname.
When the user scale is small, \othersys{IBMHom} is faster.
However, in real-world applications, such a small scale of users is almost impossible.
For \othersys{FPE}, although its communication overhead and user runtime are less than \sysname, its server has to conduct double times modular exponentiation to unmask the encrypted aggregation result, which greatly increases its runtime.
Therefore, when applied to federated learning, \sysname is still more practical than \othersys{FPE}.

\section{Related Work}
\label{sec_relatedwork}
Google's federated learning is a kind of privacy-preserving machine learning framework originally proposed for the mobile crowdsensing scenario~\cite{mcmahan2016communication}.
Due to the high performance on security and efficiency, federated learning attracts a lot of attention as soon as being proposed.
Up to now, most of the existing federated learning schemes are designed towards the stochastic gradient descent (SGD) based neural networks.
For example, Wang \textit{et al.}~\cite{wang2018edge} provided an edge computing based federated learning scheme for the convolutional neural network (CNN) in the Internet of Things (IoT) environment.
Mcmahan \textit{et al.}~\cite{mcmahan2017learning} applied federated learning to the long-short term memory network (LSTM) based language model and gain better performance than the traditional centralized machine learning method.
Smith \textit{et al.}~\cite{smith2017federated} proposed a general federated learning framework for the neural network to simultaneously process multi-tasks, which solves the stragglers and fault tolerance problems in the real-world network and significantly improved the efficiency of the original federated learning framework.
Nonetheless, there is none of the existing work that gives a systematical federated learning scheme for XGBoost, a special tree structure machine learning model.

To avoid the adversary to analyze the hidden information about private user data from the uploaded gradient values~\cite{wang2019beyond}, almost all of the current federated learning schemes introduce the secure aggregation mechanism.
The existing secure aggregation schemes for federated learning mainly depend on three types of cryptographic tools.
The first and most popular tool is differential privacy (DP).
Mcmahan \textit{et al.} \cite{mcmahan2017learning} is one of the outstanding works using DP to protect the gradient's security.
Nonetheless, the introduction of noises for DP is pointed to be able to lead non-erasable accuracy loss to the trained model~\cite{chen2016oblivious}.
The second tool is secret sharing (SS), especially the Shamir's secret sharing scheme.
In~\cite{bonawitz2017practical}, Bonawitz proposed a novel SS based secure aggregation scheme against user dropout.
However, since having to operate data reconstruction  for all users, no matter the user is dropout or not, the communication cost of \cite{bonawitz2017practical} explodes with the increasing  of the user number.
The last tool is homomorphic encryption (HE).
For HE, the most commonly used HE method is the Pailliar cryptosystem~\cite{paillier1999public} and its variants~\cite{bresson2003simple, Damg2001thresh}.
Although many HE schemes are proposed~\cite{truex2018hybrid, xu2019hybridalpha, yang2019quas}, none of them solve the forced aggregation problem while preserving the robustness against user dropout.

\section{Conclusion} 
\label{sec_conclusion}
In this paper, we proposed a privacy-preserving federated extreme gradient boosting scheme (\sysname) for mobile crowdsensing.
In \sysname, a new hybrid secure aggregation scheme is first presented by combining homomorphic encryption and secret sharing, which can force the central server to conduct the aggregation operation, and is robust against user dropout.
Then, using the newly designed secure aggregation scheme, we designed a suite of secure protocols to implement the classification and regression tree building of XGBoost.
Comprehensive experiments were conducted to evaluate the effectiveness and efficiency of \sysname. 
Experiment results showed that \sysname made it possible to train an XGBoost with negligible performance loss, and attained computation and communication cost reduction for secure aggregation.

\ifCLASSOPTIONcompsoc
  \section*{Acknowledgments}
\else
  \section*{Acknowledgment}
\fi
This work was supported by the National Natural Science Foundation of China (Grant No. 61872283, 61702105, U1764263, U1804263), the China 111 Project (No. B16037).

\footnotesize
\bibliographystyle{IEEEtran}
\bibliography{references}

\end{document}